\documentclass[aps,prb,twocolumn,showpacs,superscriptaddress,groupedaddress]{revtex4-1}

\usepackage{graphicx}
\usepackage{epstopdf}     
\graphicspath{{../FIGS/}}

\usepackage{amssymb}
\usepackage{rotating}  

\begin{document}

\title{Valence and magnetism in $\rm EuPd_3S_4$ and $\rm (Y,La)_xEu_{1-x}Pd_3S_4$}

\date{\today}

\author{D. H. Ryan}
\affiliation{Physics Department and Centre for the Physics of Materials,
McGill University, 3600 University Street, Montreal, Quebec, H3A 2T8, Canada}

\author{Sergey L. Bud'ko}
\author{Brinda Kuthanazhi}
\author{Paul C. Canfield}
\affiliation{Ames National Laboratory, and Department of Physics and Astronomy,
Iowa State University, Ames, Iowa 50011, USA}

\begin{abstract}

$^{151}$Eu M\"ossbauer spectroscopy shows that yttrium substitution in mixed-valent
$\rm EuPd_3S_4$ drives the initial 50:50 mix of Eu$^{3+}$ and Eu$^{2+}$ towards
pure Eu$^{2+}$, whereas lanthanum substitution has the opposite effect, but only for
substitution levels above 50\%. We find that total valence electron count and
chemical pressure effects cannot account for the observed behaviour, however
conserving the cell volume provides a consistent description of the
changes in the Eu$^{2+}$:Eu$^{3+}$ ratio.
Remarkably, lanthanum substitution also leads to a clear transition from
static mixed-valent behavior at lower temperatures to dynamic mixed
valent behavior at higher temperatures, with
the onset temperature monotonically increasing with Eu content and
extrapolating to a value of $\sim$340~K for the pure $\rm EuPd_3S_4$ compound.
Magnetic order persists at least as far as x=0.875 in both series,
despite the drastic reduction in the amount of moment-carrying Eu$^{2+}$ ions.

\end{abstract}

\maketitle

\section{Introduction}

The rare earth palladium sulphides $\rm RPd_3S_4$ have been reported for
the majority of the rare earths, including yttrium \cite{keszler2369,
wakeshima1, wakeshima226, bonville263}. They all crystallise in the cubic
$\rm NaPt_3O_4$ structure ($Pm\overline{3}n$ \#223) with the rare earth occupying
the 2$a$ site forming a bcc sublattice, the palladium on the 6$d$ site and the
sulphur on the 8$e$ site. Remarkably, although the $\rm RPd_3S_4$
phases exist for the trivalent rare earths, but apparently not for the divalent
alkaline earths (Ca and Sr), when prepared with europium \cite{wakeshima117} or
ytterbium \cite{bonville263} a roughly 50:50 mix of divalent and trivalent rare
earth is found.

Here we will use chemical substitution of yttrium and lanthanum for europium to
investigate the stability of the valence distribution and its effects on
magnetic ordering. Although both $^{170}$Yb and $^{151}$Eu M\"ossbauer
spectroscopy can generally be used to identify the valence of their respective
target ions, for $^{170}$Yb M\"ossbauer spectroscopy, the isomer shift between
the two valence states is extremely small so the technique is almost totally
dependent on the presence of an electric field gradient (efg) at the Yb$^{3+}$ ions to
identify trivalent ytterbium. Unfortunately the high symmetry of the 2$a$ site
makes the efg contribution effectively zero and the presence of the Yb$^{3+}$ ions
is only apparent in the magnetically ordered state well below T$_N \sim$2~K
\cite{bonville263}. We will therefore only study $\rm EuPd_3S_4$ using
$^{151}$Eu M\"ossbauer spectroscopy where the two valence states are clearly
isolated by a large difference in isomer shift, even at ambient temperatures.

We find that, by substituting Y for Eu, the remaining Eu sites become more and
more divalent.  In contrast, by substituting La for Eu we find that, initially,
the remaining Eu sites stay roughly a 50:50 mixture of di- and tri-valent Eu,
but for higher La substitution levels the remaining Eu rapidly becomes more trivalent.
La substitution also leads to a transition from statically mixed valent
behavior at lower temperatures to dynamically mixed valent behavior at higher
temperatures with the onset temperature (T$_{onset}$) monotonically increasing with Eu
content and passing through room temperature as pure $\rm EuPd_3S_4$ is approached.
Despite the decreasing fraction of moment-carrying Eu$^{2+}$ ions, 
both $\rm Y_xEu_{1-x}Pd_3S_4$ and $\rm La_xEu_{1-x}Pd_3S_4$ continue to exhibit
some form of magnetic order at least as far as x=0.875, with transition
temperatures of $\sim$3~K (Y) and $\sim$6~K (La).

\section{Experimental methods}

Polycrystalline samples of $\rm EuPd_3S_4$ and $\rm
(Y,La)_xEu_{1-x}Pd_3S_4$ were prepared from stoichiometric mixtures of EuS
(99.9\% -- American Elements) $\rm Y_2S_3$ (99.9\%),  $\rm La_2S_3$ (99\%), Pd
(99.95\%) and S (99.5\%), all from Alfa-Aesar. The powders were mixed and then
pressed to form a dense pellet. This was loaded into an alumina crucible and
sealed under a partial pressure of helium in a fused silica tube. The sample was
heated to 650$^{\circ}$C over three hours, held for an hour and then taken to
900$^{\circ}$C over a further three hours. After 75 hours at 900$^{\circ}$C the
sample was furnace cooled and removed once it reached ambient temperature.
In most cases this single thermal cycle was found to yield a single-phased
product, however when an impurity was found (typically PdS seen by x-ray
diffraction or EuS seen in susceptibility vs. temperature) the sample was
crushed, pressed and subjected to a second 75~hr annealing cycle to
900$^{\circ}$C to remove the impurity.

X-ray diffraction measurements were made on a Rigaku Miniflex-II diffractometer
using a Cu-K$_{\alpha}$ source. The instrument calibration was checked using NIST
676a $\rm Al_2O_3$ and found to be consistent within fitted uncertainties. Full
Rietveld refinement of the diffraction pattens was carried out using the
GSAS/EXPGUI packages \cite{GSAS, EXPGUI}. 
As all three species occupy special sites in the $Pm\overline{3}n$ structure, no
positional parameters were adjusted during the fits.  For the yttrium and
lanthanum substituted samples only random occupation of the Eu(2$a$) site was
considered.

$^{151}$Eu M\"ossbauer spectroscopy measurements were carried out using a
4~GBq $^{151}$SmF$_3$ source, driven in sinusoidal mode.  The drive motion was
calibrated using a standard $^{57}$Co\underline{Rh}/$\alpha$-Fe foil. Isomer
shifts are quoted relative to EuF$_3$ at ambient temperature.  The 21.6~keV
gamma rays were recorded using a thin NaI scintillation detector.  For
temperatures above 5~K, the samples were cooled in a vibration-isolated
closed-cycle helium refrigerator with the sample in a helium exchange gas.
Temperatures below 5~K were achieved using a helium flow cryostat while pumping
on the sample space and using a needle valve to throttle the flow. The spectra
were fitted to a sum of Lorentzian lines with the positions and intensities
derived from a full solution to the nuclear Hamiltonian \cite{voyer}.

Temperature- and magnetic field- dependent magnetization measurements ($\rm
1.8~K \leq T \leq 300~K$, $\rm 0~T \leq \mu_0H \leq 7~T$) were performed using a
Quantum Design MPMS-3 SQUID magnetometer. The sample was confined in a \#4
gelatin capsule and a transparent drinking straw was used as a sample holder.
Low-temperature heat capacity measurements were made using semiadiabatic thermal
relaxation technique as implemented in the heat capacity option of a Quantum
Design Physical Property Measurement System (PPMS). For selected samples, the
$^3$He option was used to cool to $\sim$0.4~K. Sintered samples of 20--70~mg
mass with at least one flat surface were mounted on a micro-calorimeter
platform using Apiezon N grease. The addenda (platform + grease) heat capacity
was measured separately for each sample and subtracted from the total heat capacity
using the PPMS software. Although the samples possibly had
reduced density, the measured sample coupling parameter took reasonable values
of more than 97\%.

\section{Results}

\subsection{$\rm EuPd_3S_4$}

%%%%%%%%%%%%%%%%%%%
\begin{figure}
\includegraphics[width=8cm]{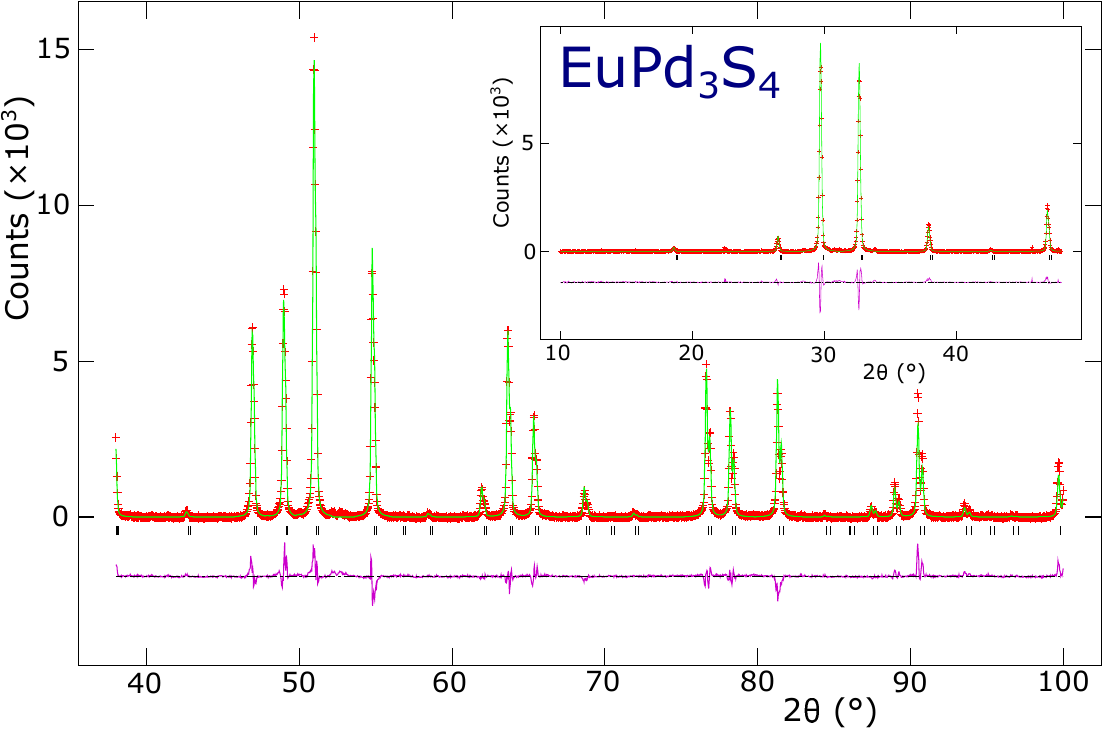}
\caption{Cu-K$_{\alpha}$ x-ray diffraction pattern for $\rm EuPd_3S_4$. Solid
line is a full Rietveld refinement using the GSAS/EXPGIU packages \cite{GSAS,
EXPGUI}. The line below the data shows the residuals. Tick marks between the
data and residual lines show the calculated positions of the Bragg peaks. The
inset shows the low-angle range. The diffraction pattern was collected in two
overlapping blocks with a longer counting time at higher angles to compensate
for the loss of intensity due to the effects of the x-ray form factor. }
\label{fig:Eu-XRD}
\end{figure}
%%%%%%%%%%%%%%%%%%%

\begin{figure}
\includegraphics[width=7cm]{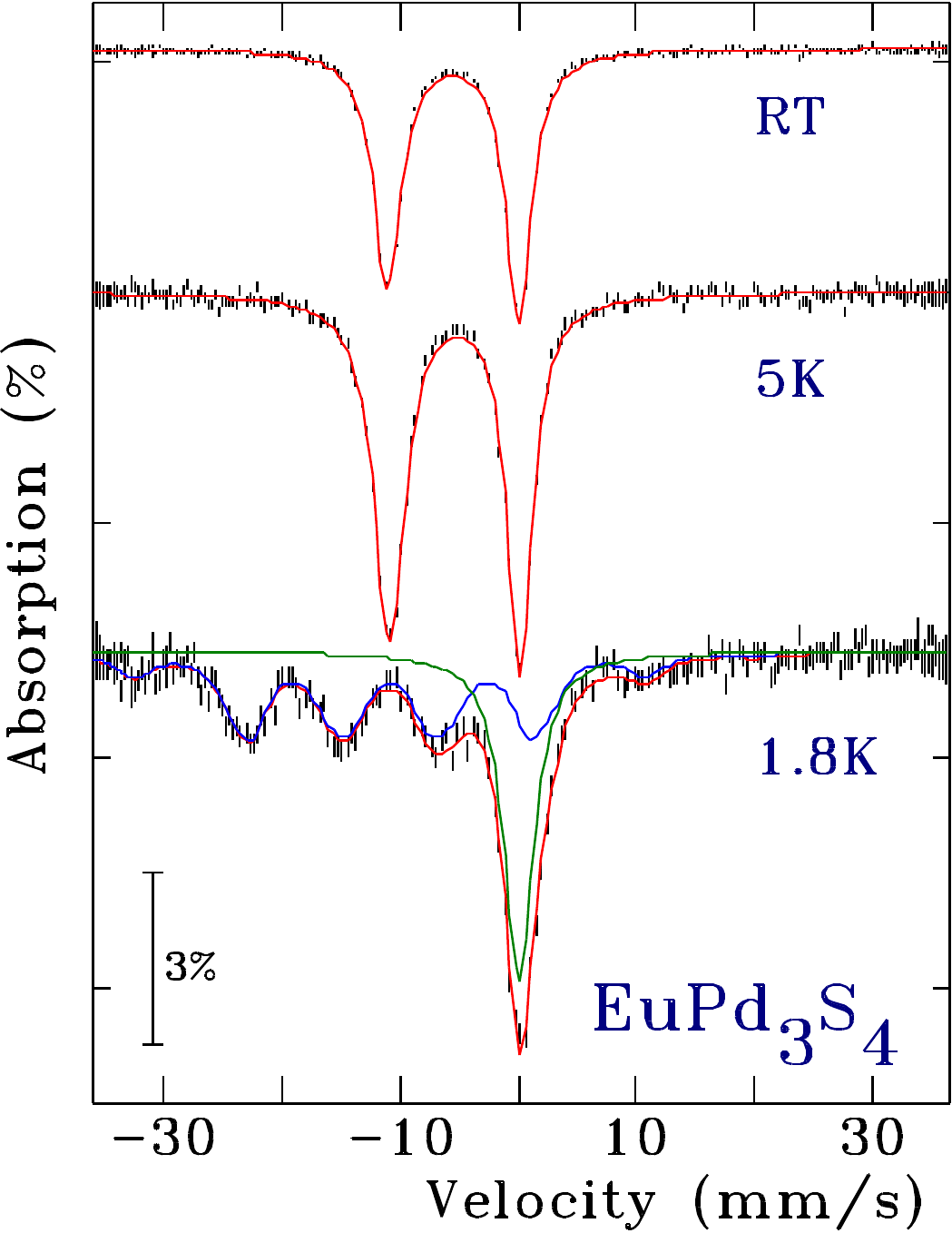}
\caption{$^{151}$Eu M\"ossbauer spectra of $\rm EuPd_3S_4$ at RT (top), 5~K
(middle), and 1.8~K (bottom). Between RT and 5K the primary changes are a
significant increase in the absorption due to conventional thermal effects and a
slight change in the 2+:3+ area ratio due to differences in the Debye
temperatures of the two components. At 5~K the two valence contributions are
almost equal in area and fully resolved, with the Eu$^{2+}$ at $-10.91$(2)~mm/s
and Eu$^{3+}$ at $+0.07$(1)~mm/s. At 1.8~K only the Eu$^{2+}$ component is
ordered and shows a hyperfine field (B$_{hf}$) of 29.9(2)~T, while the Eu$^{3+}$
component is unchanged. The solid red lines are fits as desribed in the text.
For the 1.8~K spectrum we also show the magnetically split Eu$^{2+}$ and
unchanged Eu$^{3+}$ components.}
\label{fig:Eu-spectra}
\end{figure}

\begin{figure}
\includegraphics[width=7cm]{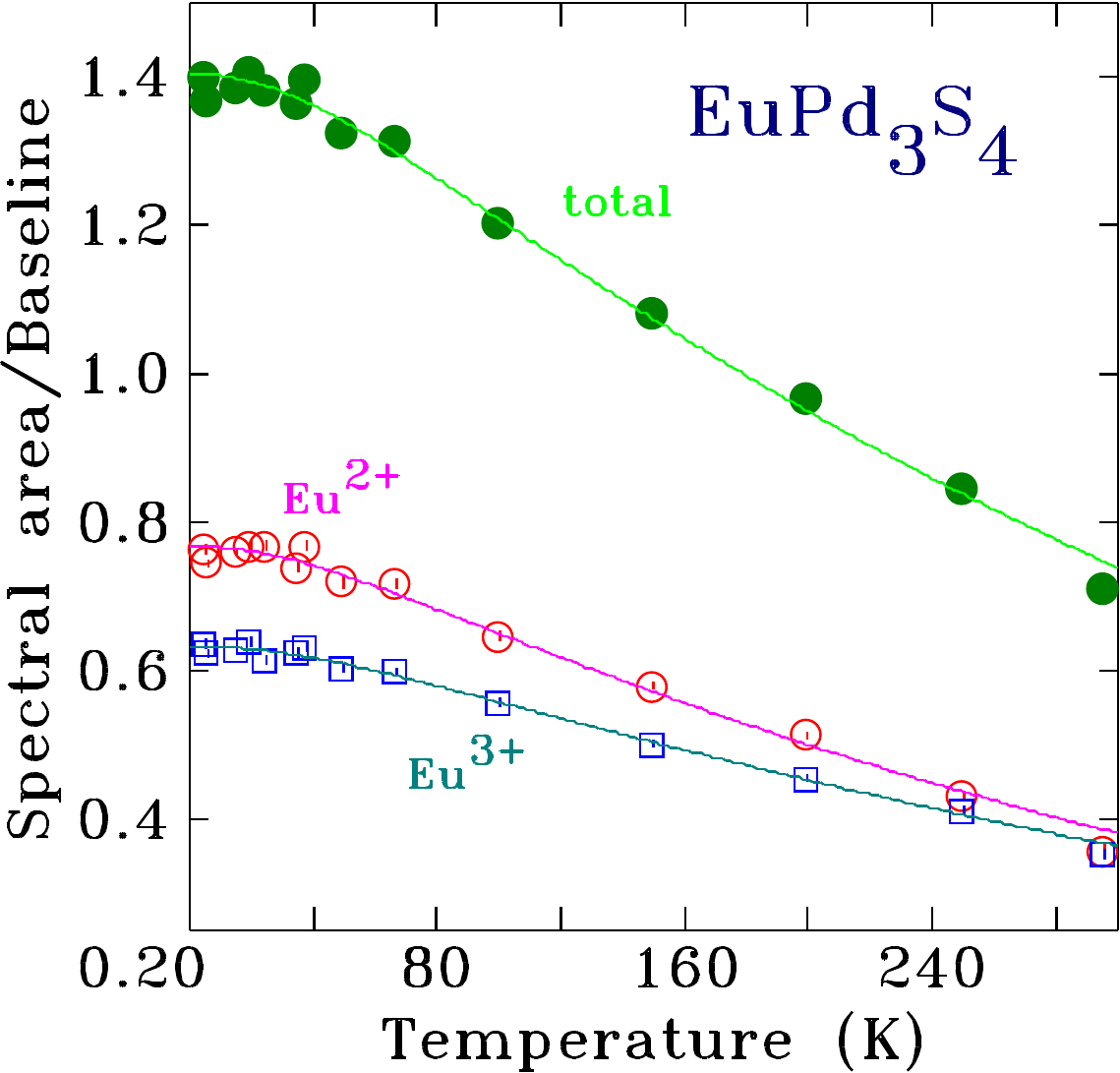}
\caption{Temperature dependence of the normalised (adjusted for total counting
time) area of both the total spectrum and the two valence components in the
$^{151}$Eu M\"ossbauer spectra of $\rm EuPd_3S_4$. Solid lines in each case are
fits to a simple Debye model yielding Debye temperatures of 227(3)~K (Eu$^{3+}$)
and 204(3)~K (Eu$^{2+}$). Fitting the total spectral area yields an average
Debye temperature of 212(3)~K.}
\label{fig:EuDebye}
\end{figure}

Fitting the x-ray diffraction pattern of the $\rm EuPd_3S_4$ sample showed it to
be single phased with the expected cubic $\rm NaPt_3O_4$ structure
\cite{wakeshima1} and a lattice parameter of $a=6.6786(1)$~\AA\ (Fig.~\ref{fig:Eu-XRD}).
The room temperature $^{151}$Eu M\"ossbauer spectrum showed two distinct
contributions from Eu$^{3+}$ and Eu$^{2+}$ in the ratio 50.4(4):49.6(4), with
the linewidth of the  Eu$^{2+}$ component being slightly broader, in
complete agreement with previous reports \cite{wakeshima117}. On cooling to 5~K
the ratio becomes 46.2(5):53.8(5) (Fig.~\ref{fig:Eu-spectra}) as the Debye
temperature of the Eu$^{3+}$ component is slightly higher than that of the Eu$^{2+}$
component. Fitting the temperature dependence of the two component areas to a
simple Debye model, as shown in Fig.~\ref{fig:EuDebye}, yields Debye temperatures
of 227(3)~K (Eu$^{3+}$) and 204(3)~K (Eu$^{2+}$). It is important to emphasise
that any {\em apparent} changes in the Eu$^{2+}$:Eu$^{3+}$ ratio with temperature in
Fig.~\ref{fig:EuDebye} do not reflect {\em actual} changes in the ratio, rather
they are the result of the different temperature dependences of the
recoil-free-fractions (often denoted {\em f}) for the two species. In order to
minimise the impacts of this effect, all valence ratios will be taken from
low-temperature spectra.

%%%%%%%%%%%%%%%%%%%
%%%%%%%%%%%%%%%%%%%

%%%%%%%%%%%%%%%%%%%
%%%%%%%%%%%%%%%%%%%

\begin{figure}
\includegraphics[width=7cm]{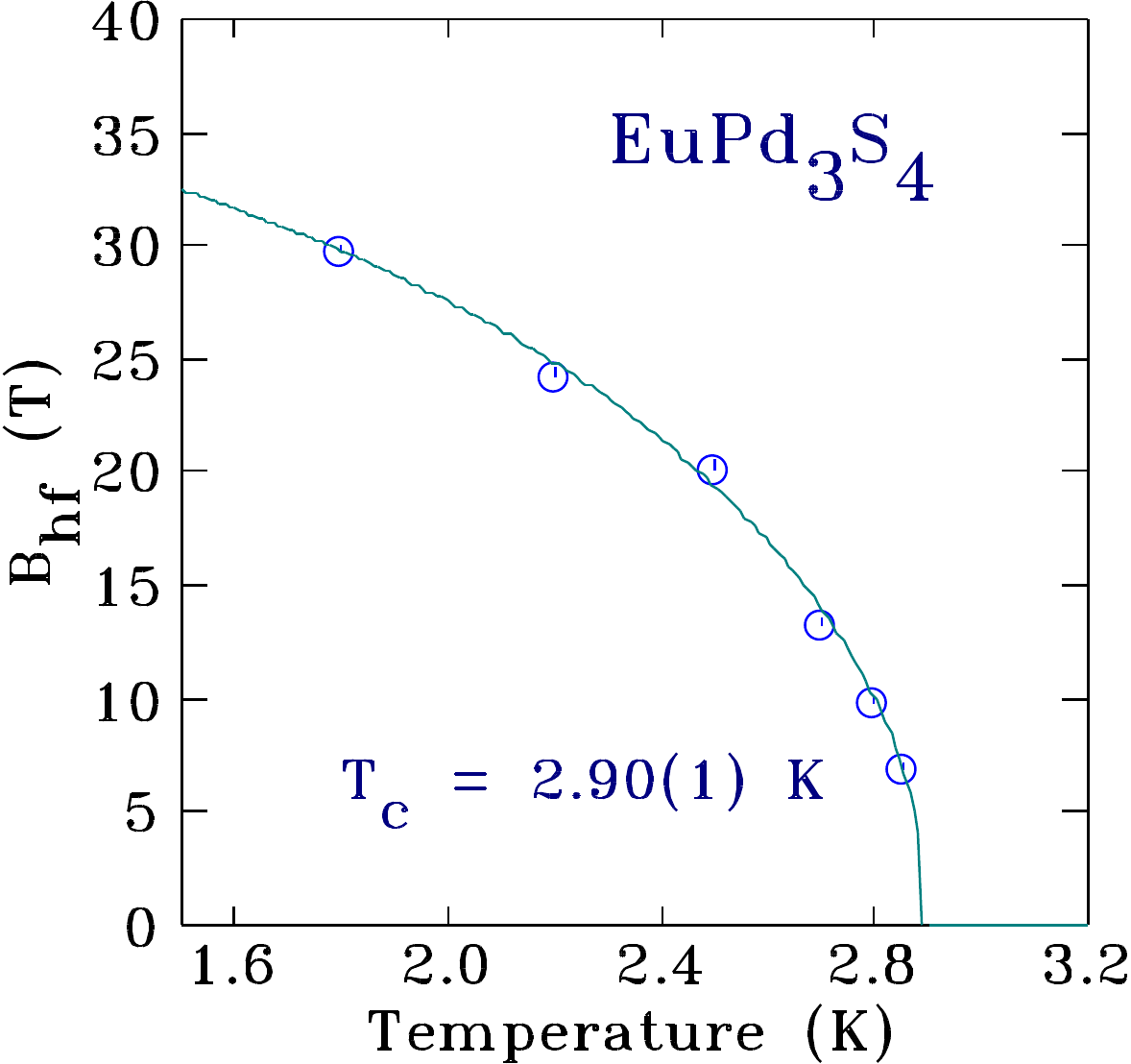}
\caption{Temperature dependence of hyperfine field (B$_{hf}$) for $\rm
EuPd_3S_4$ fitted using the expected $J = \frac{7}{2}$ Brillouin function to
obtain the ordering temperature of T$_N$=2.90(1)~K.  }
\label{fig:Eu-Tdep}
\end{figure}

\begin{figure}
\includegraphics[width=7cm]{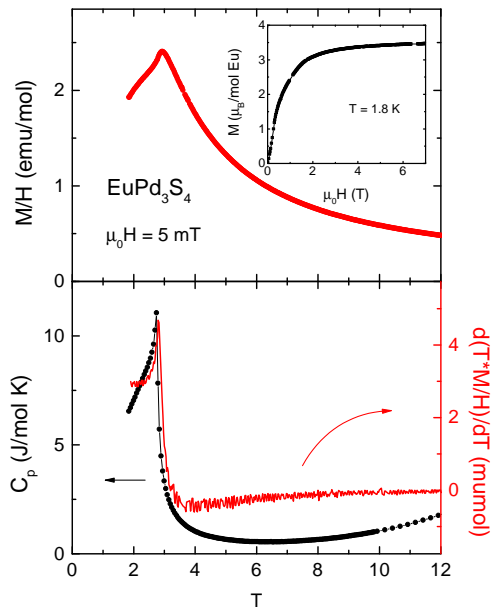}
\caption{Top: DC susceptibility vs. temperature for $\rm EuPd_3S_4$ showing a cusp
at T$_N$$\sim$3~K. Inset shows M {\it vs.} $\mu_0$H at 1.8~K confirming that
half of the europium is divalent.
Bottom: Temperature dependence of the heat capacity (black points) revealing
the corresponding cusp associated with the AF transition, with d(T$\cdot$M/H)/dT
(red line) showing that C$_p$(T) and the temperature derivative of T$\cdot$M/H
take the same form around the transition.}
\label{fig:Eu-Cp-chi}
\end{figure}

On cooling to 1.8~K, the Eu$^{2+}$ component develops a clear magnetic splitting
whereas the Eu$^{3+}$ component is unchanged, consistent with trivalent europium
being non-magnetic (Fig.~\ref{fig:Eu-spectra}). In principle one might
anticipate a small transferred field at the Eu$^{3+}$ sites from the ordered
Eu$^{2+}$ moments, however none was observed and no increase in the width of the
Eu$^{3+}$ component was detected (B$_{hf} \ll$1~T), perhaps as a result of
cancellations arising from the antiferromagnetic ordering of the Eu$^{2+}$
moments.
The hyperfine field (B$_{hf}$) for the Eu$^{2+}$ component at 1.8~K is
29.9(2)~T, typical for ordered Eu$^{2+}$, and fitting the temperature dependence
shown in Fig.~\ref{fig:Eu-Tdep} to the expected $J = \frac{7}{2}$ Brillouin
function yields a T=0 B$_{hf}$=37.9(6)~T and an ordering temperature of
T$_N$=2.90(1)~K in good agreement with both previous work \cite{wakeshima117}
and our own susceptibility and C$_p$ data (Fig.~\ref{fig:Eu-Cp-chi}). The clear
cusp in the susceptibility
{\it vs.} temperature is consistent with antiferromagnetic (AF) ordering and it is
accompanied by a sharp peak in the heat capacity. The high-field magnetisation
curve taken at 1.8~K (inset to Fig.~\ref{fig:Eu-Cp-chi}) shows that the system is readily
saturated despite the AF order, consistent with the low ordering temperature as
well as the low anisotropy typically
associated with the Eu$^{2+}$ ion. Furthermore, the maximum moment observed in
the applied field of 7~T suggests a Eu$^{2+}$ fraction of 50\% (assuming a
moment of 7~$\mu_B$/Eu$^{2+}$) consistent with the 53.8(5)\% derived above from the
$^{151}$Eu M\"ossbauer spectrum at 5~K, and with earlier results \cite{Abe5366,
wakeshima117}.

%%%%%%%%%%%%%%%%%%%
%%%%%%%%%%%%%%%%%%%

%%%%%%%%%%%%%%%%%%%
%%%%%%%%%%%%%%%%%%%

The unusual and apparently stable valence mix in $\rm EuPd_3S_4$ leads to the
question: ``why?''. What makes europium (and ytterbium) ``special''? How robust
is the valence distribution? Can we change it?

As the Eu$^{3+}$ ion is smaller than the Eu$^{2+}$ ion, one might
expect hydrostatic pressure to promote Eu$^{2+}$~$\rightarrow$~Eu$^{3+}$
conversion. Alternatively, if we force some fraction of the R sites to be
unambiguously trivalent, by replacing some of the europium with a formally
trivalent ion, will this cause more of the remaining europium to become divalent
to preserve the average electron count?

Although the driving that can be achieved by chemical substitution is not as
clean as that generated by direct hydrostatic pressure, it is much easier to
make direct measurements of the valence distribution, magnetisation and
transition temperatures in doped samples at ambient pressures. We turn therefore
to an investigation of the impacts of chemical substitution on $\rm EuPd_3S_4$
using the non-moment bearing, trivalent La and Y ions with $r_{ionic} \rm
(Eu^{2+}$) $\gtrsim$ $r_{ionic} \rm (La^{3+}$) $>$ $r_{ionic} \rm (Eu^{3+}$) $>$
$r_{ionic} \rm (Y^{3+}$). If, on the one hand, the total valence electron count
is a dominant factor, then partially replacing the europium with an
unambiguously trivalent ion should lead to a compensating increase in the Eu$^{2+}$
fraction. On the other hand, if preserving the cell volume is critical, then the
lattice expansion that would be caused by introducing the $r_{ionic} \rm
(La^{3+}$) $\lesssim$ $r_{ionic} \rm (Eu^{2+}$) lanthanum ion could be
compensated by some Eu$^{2+}$~$\rightarrow$~Eu$^{3+}$ conversion (the reverse
being expected for the $r_{ionic} \rm (Y^{3+}$) $<$ $r_{ionic} \rm (Eu^{3+}$)
yttrium ion). Alternatively, if the substitutions act as chemical pressures, then
expanding the cell using lanthanum substitution should lead to
Eu$^{3+}$~$\rightarrow$~Eu$^{2+}$ conversion (again, the reverse process being
expected for yttrium substitution). As we show below, we can clearly distinguish
between these three options.

\subsection{Valence impacts}

%%%%%%%%%%%%%%%%%%%
\begin{figure}
\includegraphics[width=8cm]{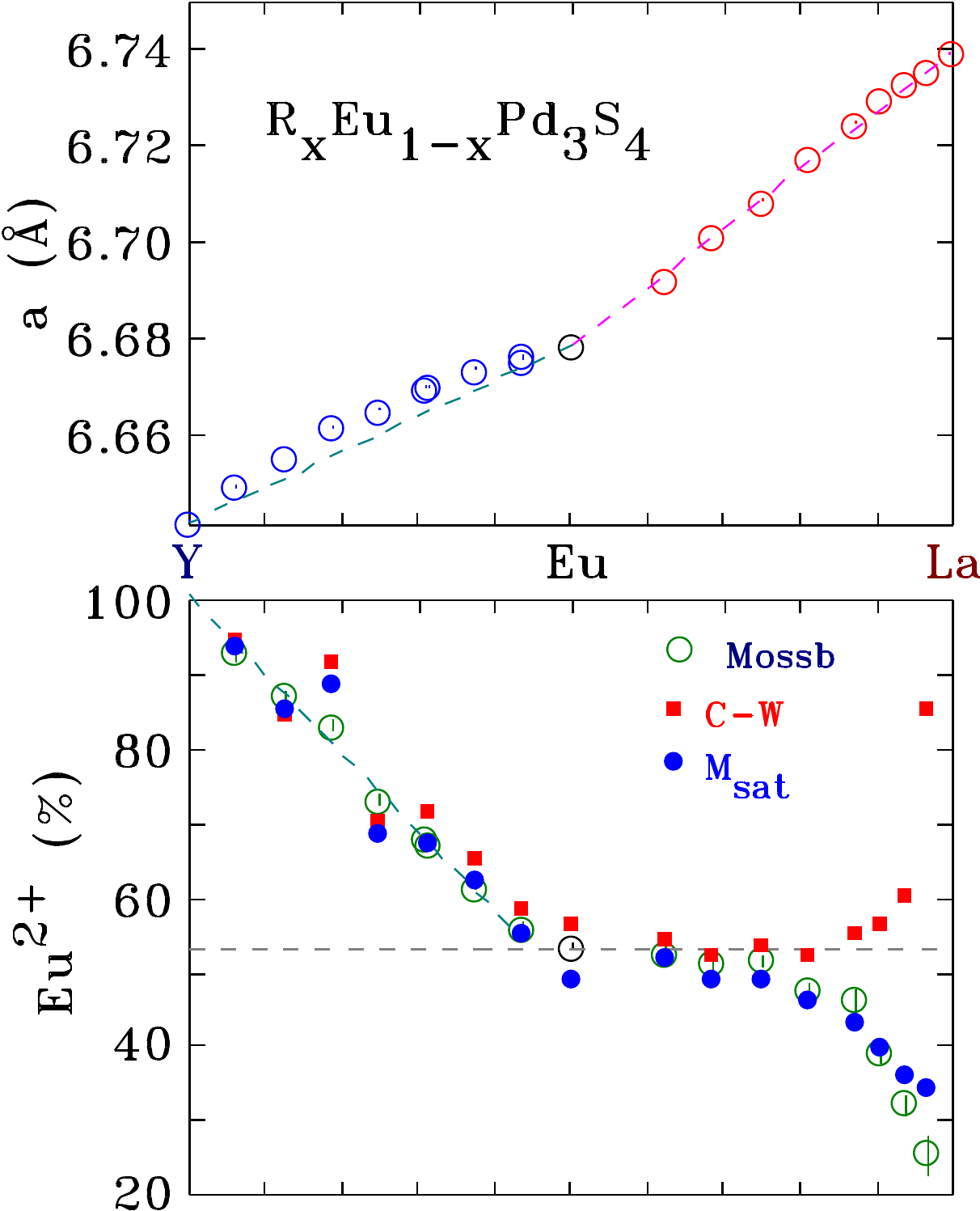}
\caption{(top) Lattice parameters for $\rm (Y,La)_xEu_{1-x}Pd_3S_4$ showing the
expansion associated with La-substitution and the contraction when yttrium is
substituted. We note that whereas the values for the
La-substituted series lie on the line connecting the pure La and Eu compounds, those
for the Y-substituted samples lie significantly above the corresponding line between
the pure Eu and Y compounds.\\
(bottom) The Eu$^{2+}$ fractions in the pure and Y,La-substituted compounds showing
that Y-substitution leads to a significant shift towards more Eu$^{2+}$, while
La-substitution initially appears to have no effect, but for $x>0.5$ there is a marked
shift towards Eu$^{3+}$. Open circles show values taken from $^{151}$Eu
M\"ossbauer spectroscopy at 5~K(Y-substituted) or 10~K(La-substituted); solid symbols show
values derived from bulk magnetisation data: high-temperature Curie-Weiss fits
(red squares), saturation magnetisation (blue circles).
The deviation of the Curie-Weiss values for the La-rich compounds reflects the
development of dynamic effects discussed in the text. }
\label{fig:lattice}
\end{figure} 
%%%%%%%%%%%%%%%%%%%

Fitting the x-ray diffraction data for the yttrium and lanthanum substituted
compounds showed that they all retained the expected $\rm NaPt_3O_4$ structure
but with progressively smaller (Y) or larger (La) lattice parameters
(Fig.~\ref{fig:lattice}). There is a clear, and significant, change in lattice
parameters for both Y and La substitution.  We detect no indication of phase
separation in the powder x-ray diffraction data, {\it i.e.} no broadening or
splitting of peaks that would suggest segregation of the samples into Eu-richer
and Eu-poorer phases.

\begin{figure}
\includegraphics[width=7cm]{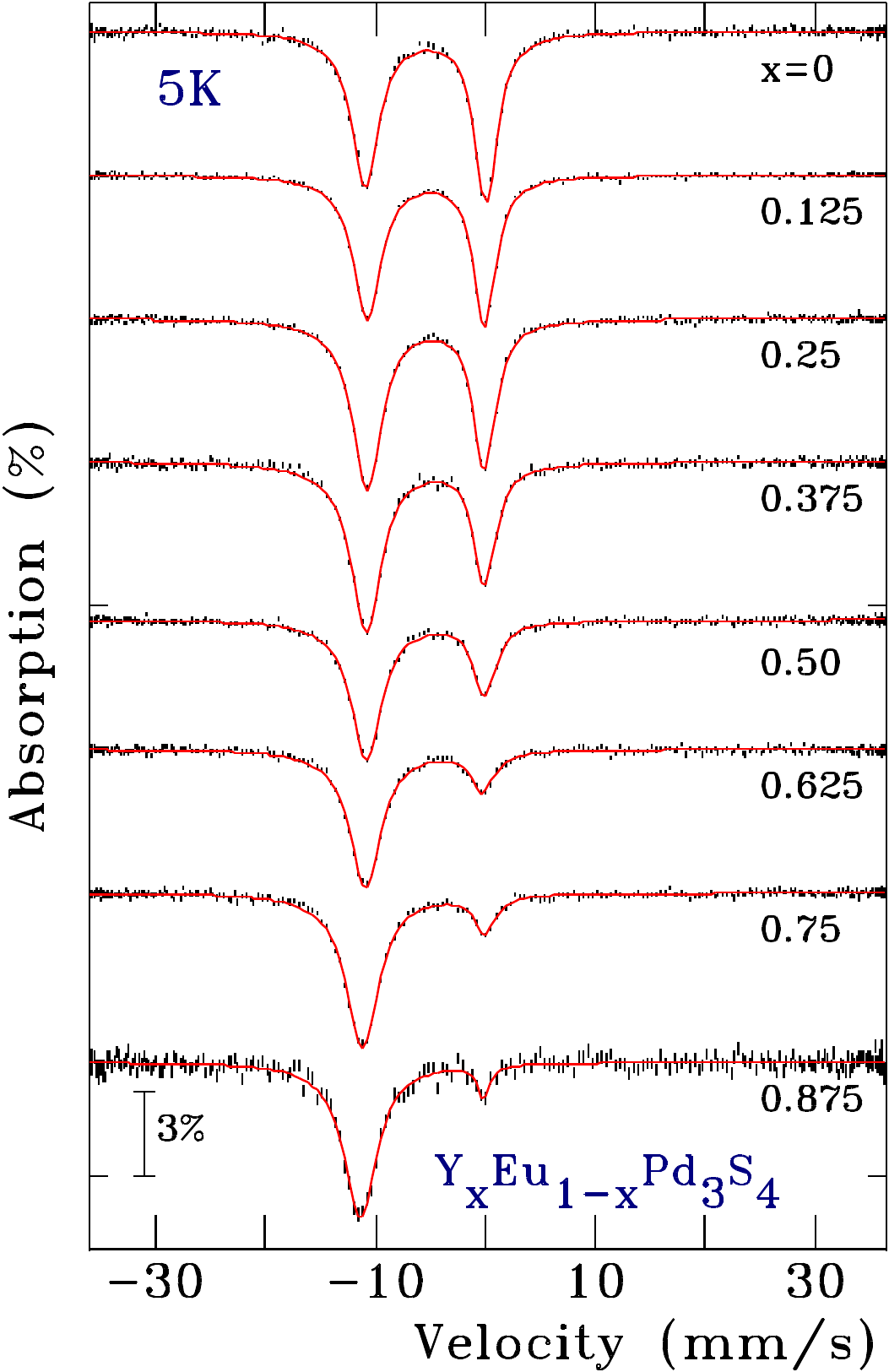}
\caption{$^{151}$Eu M\"ossbauer spectra of $\rm Y_xEu_{1-x}Pd_3S_4$ at 5~K
showing the loss of the Eu$^{3+}$ component (the line at $\sim$0~mm/s) as the
level of yttrium substitution increases. The spectrum for x=0.875 has been rescaled by
a factor of 2.5 to compensate for the low Eu content.}
\label{fig:Y-spectra}
\end{figure}

\begin{figure}
\includegraphics[width=10cm]{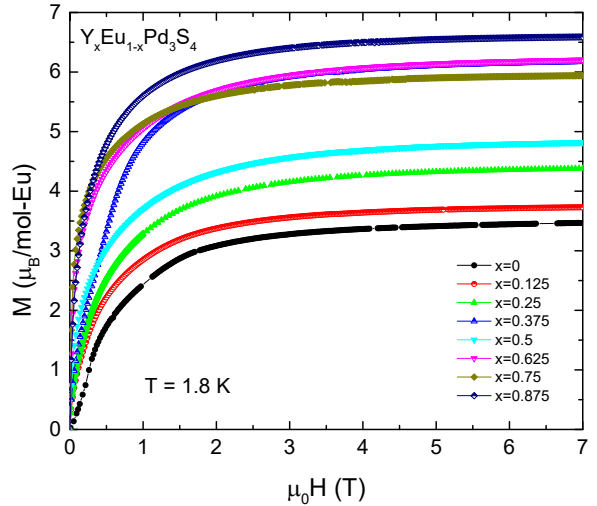}
\caption{High-field magnetisation curves measured at 1.8~K for $\rm
Y_xEu_{1-x}Pd_3S_4$ showing the increase in moment/mol-Eu with increasing
yttrium content as the substitution promotes Eu$^{3+}$~$\rightarrow$~Eu$^{2+}$
conversion. All of the curves appear to saturate in
about 3--4~T, reflecting the low anisotropy associated with the Eu$^{2+}$ ion. }
\label{fig:Y-magn}
\end{figure}

For the Y-substituted series, the
fitted lattice parameters all lie visibly above the line connecting the Eu and Y
compounds, suggesting that a shift towards more of the larger Eu$^{2+}$ ion
occurs, as the yttrium content is increased. This valence shift is confirmed
directly by the 5~K $^{151}$Eu M\"ossbauer spectra shown in
Fig.~\ref{fig:Y-spectra}, where the line near 0~mm/s associated with the
trivalent europium decreases rapidly in intensity as the level of yttrium
substitution increases. (All valence ratios for the Y-substituted series were
taken from 5~K spectra to minimise the impacts of {\em f}-factor differences
for the two species, as noted above.) Both magnetisation measurements in the
ordered state at 1.8~K (Fig.~\ref{fig:Y-magn}) and Curie-Weiss fits to the
temperature dependence of the susceptibility above 10~K, further support these
observations (see below). The Eu$^{2+}$ fractions derived from all three
measurements are summarised in the lower panel of Fig.~\ref{fig:lattice}.

%%%%%%%%%%%%%%%%%%%
%%%%%%%%%%%%%%%%%%%

%%%%%%%%%%%%%%%%%%%
%%%%%%%%%%%%%%%%%%%

\begin{figure}
\includegraphics[width=7cm]{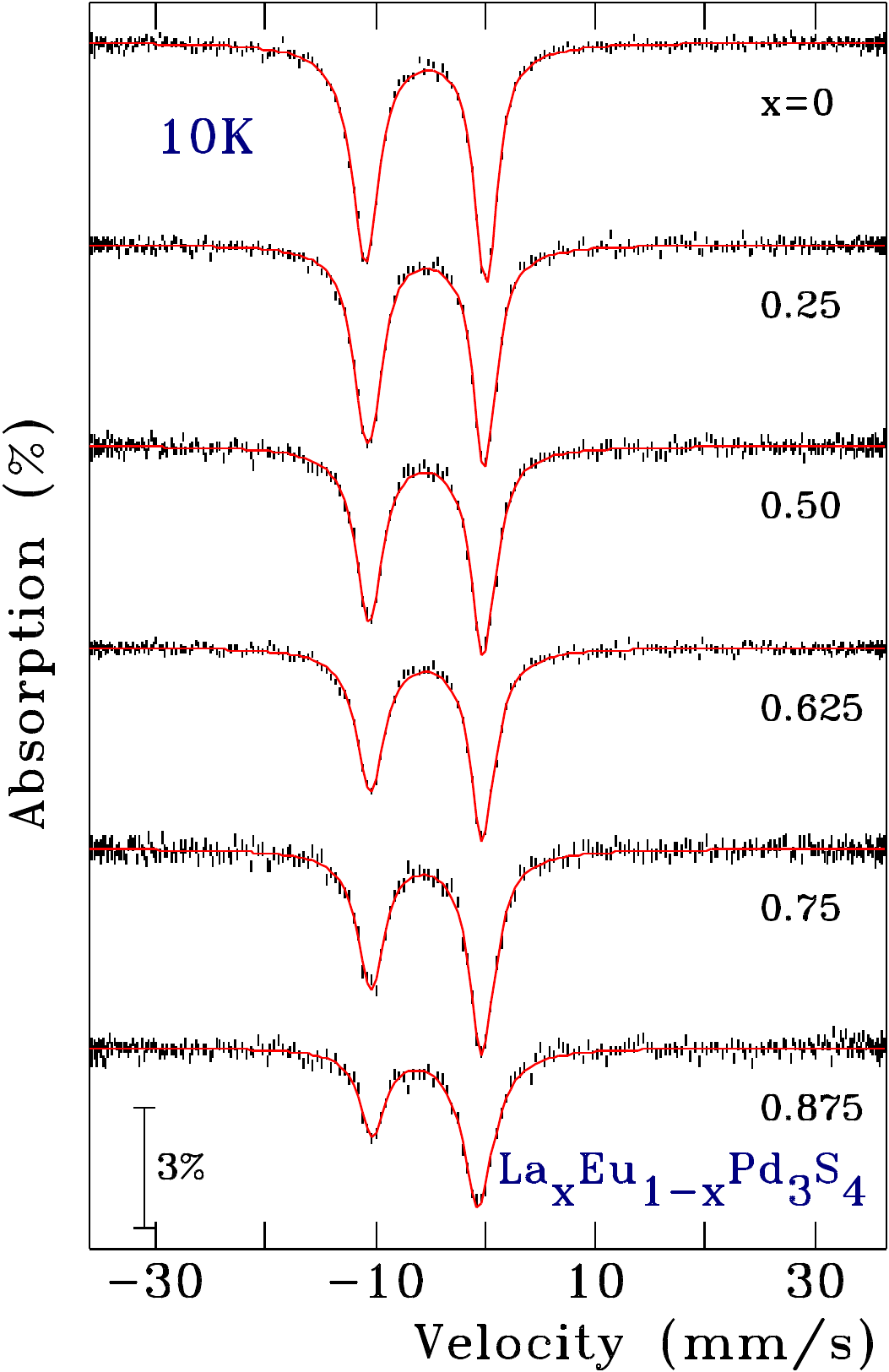}
\caption{$^{151}$Eu M\"ossbauer spectra of $\rm La_xEu_{1-x}Pd_3S_4$ at 10~K
showing an essentially constant Eu$^{2+}$:Eu$^{3+}$ distribution up to x=0.50
followed by a marked drop in the Eu$^{2+}$ fraction above that point. Spectra
taken at 10~K are used here to avoid complications associated with the magnetic
ordering of the Eu$^{2+}$ component at 6~K.}
\label{fig:La-10K-spectra}
\end{figure}

\begin{figure}
\includegraphics[width=10cm]{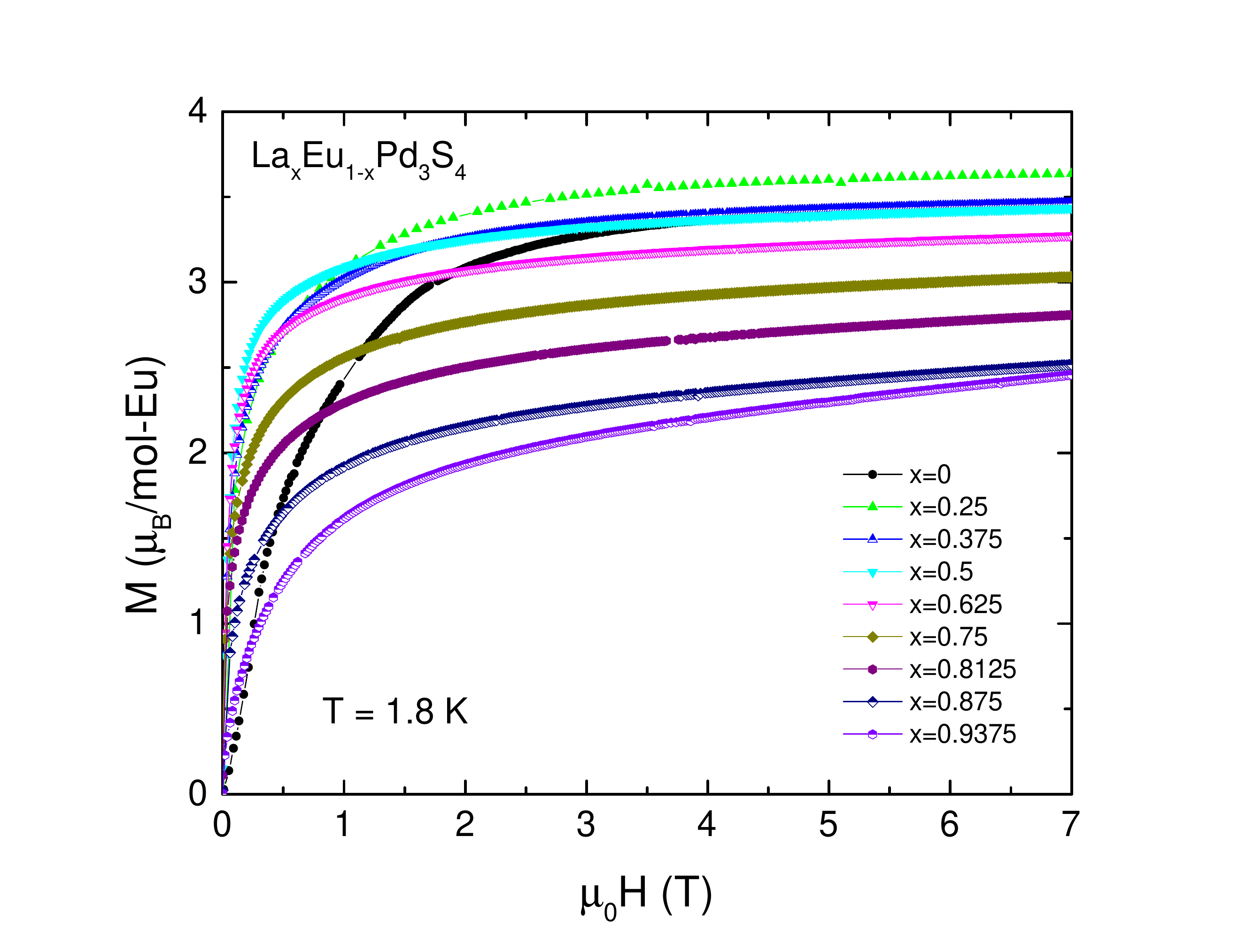}
\caption{High-field magnetisation curves measured at 1.8~K for $\rm
La_xEu_{1-x}Pd_3S_4$ showing an approximately constant moment/mol-Eu with
increasing lanthanum content up to $x=0.5$ as the Eu$^{2+}$:Eu$^{3+}$ ratio
remains largely independent of the lanthanum content, then there is a marked
drop as rapid Eu$^{2+} \rightarrow $Eu$^{3+}$ occurs. }
\label{fig:La-magn}
\end{figure}

By contrast with the yttrium-substituted series, the effects of lanthanum
substitution are more complex and nuanced.  The lattice parameters of the
La-substituted compounds lie much closer to the line connecting the two end
members (upper panel of
Fig.~\ref{fig:lattice}) for all substitution levels, and the Eu$^{2+}$ fraction
derived from $^{151}$Eu M\"ossbauer spectroscopy at 10~K (Fig.~\ref{fig:La-10K-spectra})
is essentially constant (lower panel of Fig.~\ref{fig:lattice}) for x$\leq$0.5.
This is consistent with the fact that $r_{ionic} \rm (Eu^{2+}$) $\gtrsim$ $r_{ionic}
\rm (La^{3+}$) $>$ $r_{ionic} \rm (Eu^{3+}$)
and initial La substitution does not force any significant change in the
Eu$^{2+}$:Eu$^{3+}$ ratio.
However, once about half of the europium has been replaced by lanthanum, further
substitution leads to a marked change in behaviour, again consistent with the
fact that $r_{ionic} \rm (La^{3+}$) is ultimately somewhat closer to $r_{ionic}
\rm (Eu^{2+}$) rather than $r_{ionic} \rm (Eu^{3+}$). It is readily apparent from
Fig.~\ref{fig:La-10K-spectra} that the line near $-$11~mm/s that corresponds to
Eu$^{2+}$ decreases in relative intensity and this rapid loss of the divalent
fraction is confirmed by the magnetisation measurements shown in
Fig.~\ref{fig:La-magn}. We use 10~K M\"ossbauer data for the La-substituted
series, rather than the 5~K data as was used for the Y-substituted series, to avoid
complications due to the onset of magnetic ordering (see below).

%%%%%%%%%%%%%%%%%%%
%%%%%%%%%%%%%%%%%%%

Finally, as is clear in Fig.~\ref{fig:lattice}, the Eu$^{2+}$ fraction derived
from Curie-Weiss fits (data shown below) to the temperature dependence of the susceptibility, for
$\rm La_xEu_{1-x}Pd_3S_4$ with $\rm x\gg0.5$ (measured on the same sample as part
of the run used to obtain the saturation magnetisation) appears to show the {\it opposite}
behaviour -- a marked {\it growth} in the effective moment and hence the derived Eu$^{2+}$
fraction. As we show below, this is most likely due to dynamic effects that lead to
intermediate valence behaviour of the europium ions and makes a simple Curie-Weiss analysis
intractable over the temperature range available to us.

%%%%%%%%%%%%%%%%%%%
%%%%%%%%%%%%%%%%%%%

%%%%%%%%%%%%%%%%%%%
\begin{figure}
\includegraphics[width=7cm]{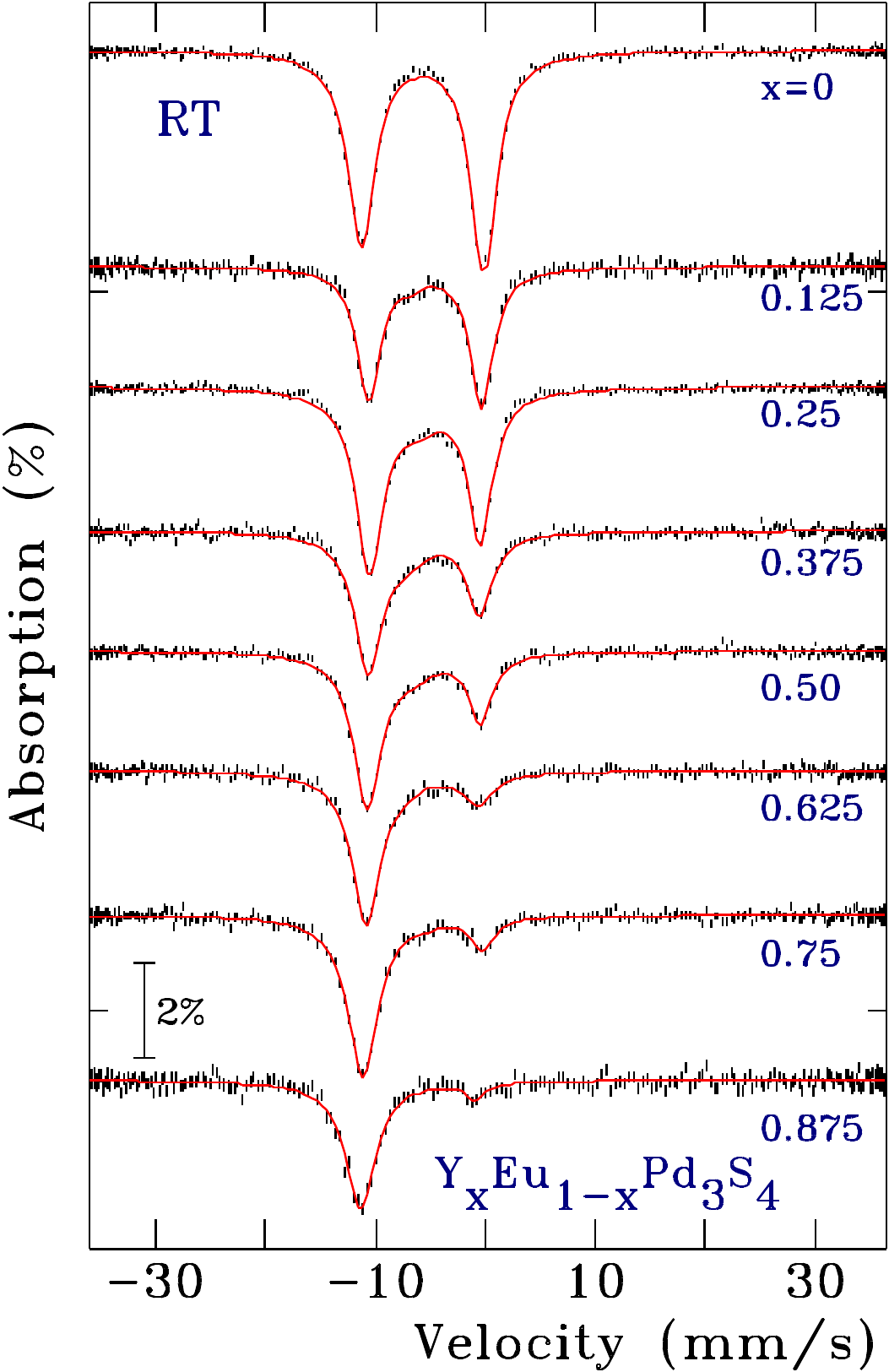}
\caption{Room temperature $^{151}$Eu M\"ossbauer spectra of $\rm Y_xEu_{1-x}Pd_3S_4$
showing minimal changes from the 5~K spectra shown in Fig.~\ref{fig:Y-spectra}
beyond the expected impacts of the differing Debye temperatures of the two
valence components.}
\label{fig:Y-RT-spectra}
\end{figure}
%%%%%%%%%%%%%%%%%%%

%%%%%%%%%%%%%%%%%%%
\begin{figure}
\includegraphics[width=7cm]{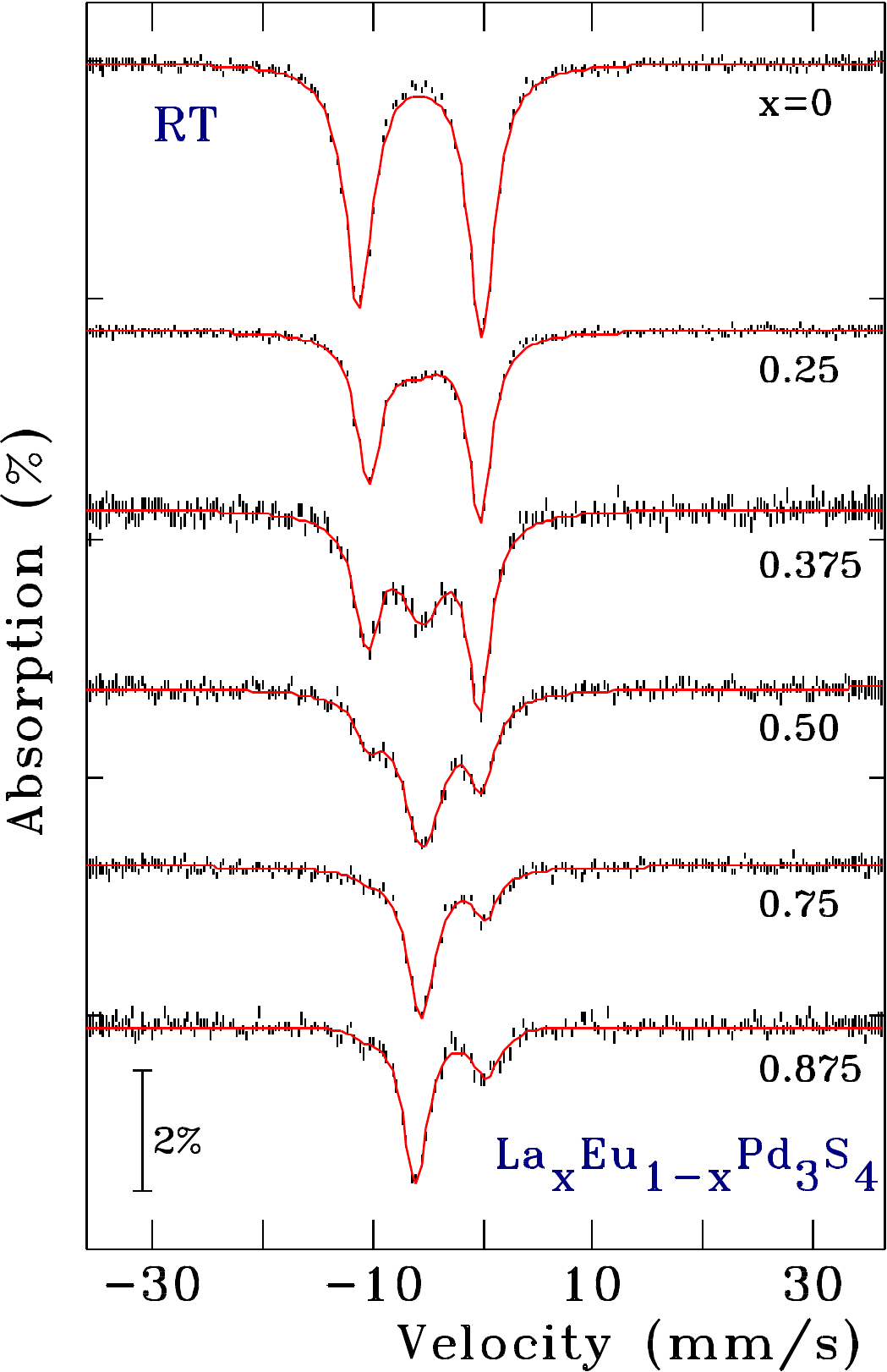}
\caption{Room temperature $^{151}$Eu M\"ossbauer spectra of $\rm La_xEu_{1-x}Pd_3S_4$
showing the development of a new, intermediate valence feature with increasing
lanthanum content.}
\label{fig:La-RT-spectra}
\end{figure}
%%%%%%%%%%%%%%%%%%%

\subsection{Dynamics in the Yttrium and Lanthanum substituted series}

%%%%%%%%%%%%%%%%%%%
\begin{figure}
\includegraphics[width=7cm]{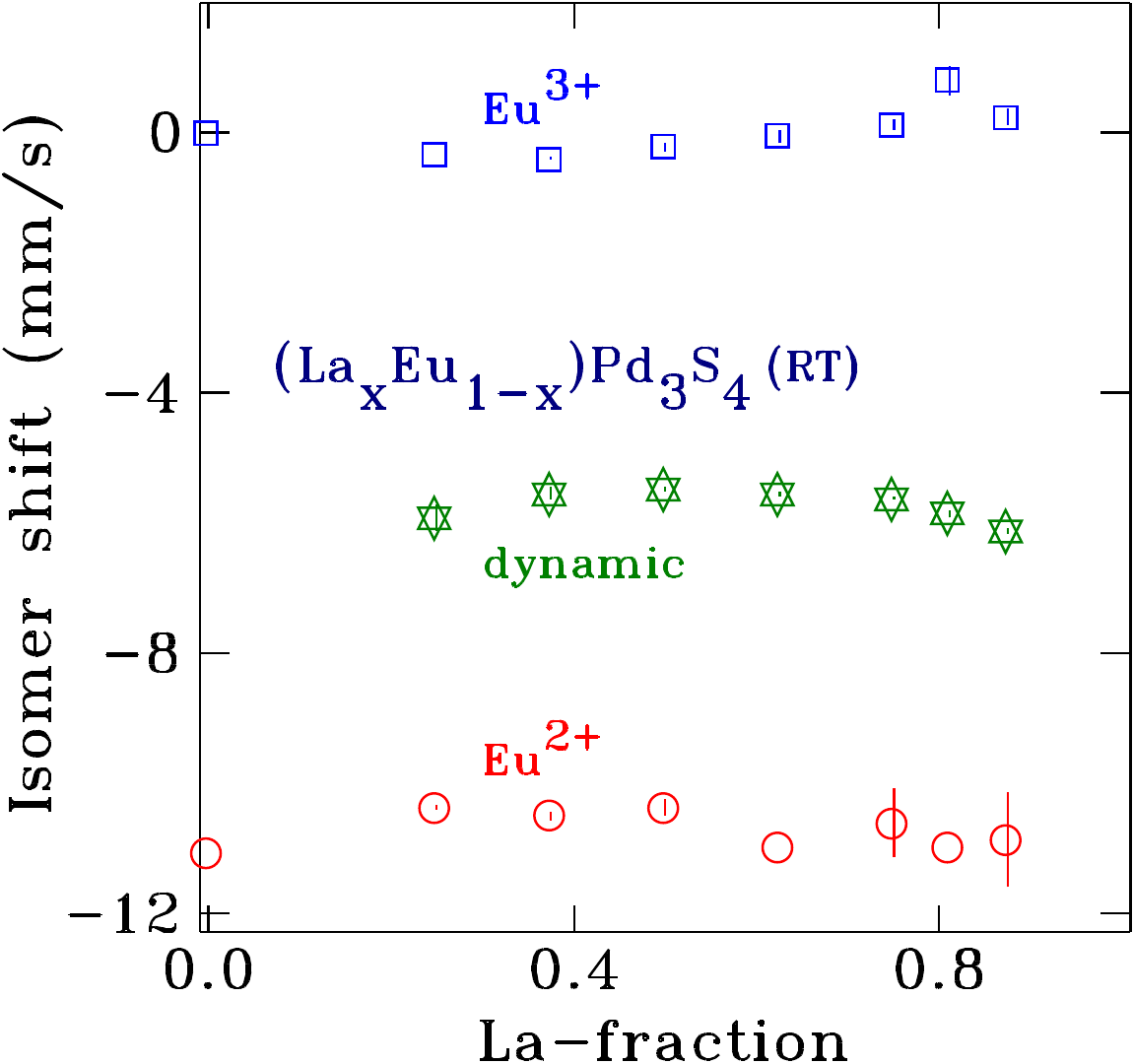}
\caption{Isomer shifts of the three components seen in the RT $^{151}$Eu
M\"ossbauer spectra of $\rm La_xEu_{1-x}Pd_3S_4$ (shown in
Fig.~\ref{fig:La-RT-spectra}) showing that the values are relatively constant
for all three, and that the dynamic, intermediate valence component lies between
the Eu$^{2+}$ and Eu$^{3+}$ component, suggesting a composition-independent
valence of $\sim$Eu$^{2.5+}$. }
\label{fig:La-RT-Relax_IS}
\end{figure}
%%%%%%%%%%%%%%%%%%%

%%%%%%%%%%%%%%%%%%%
\begin{figure}
\includegraphics[width=7cm]{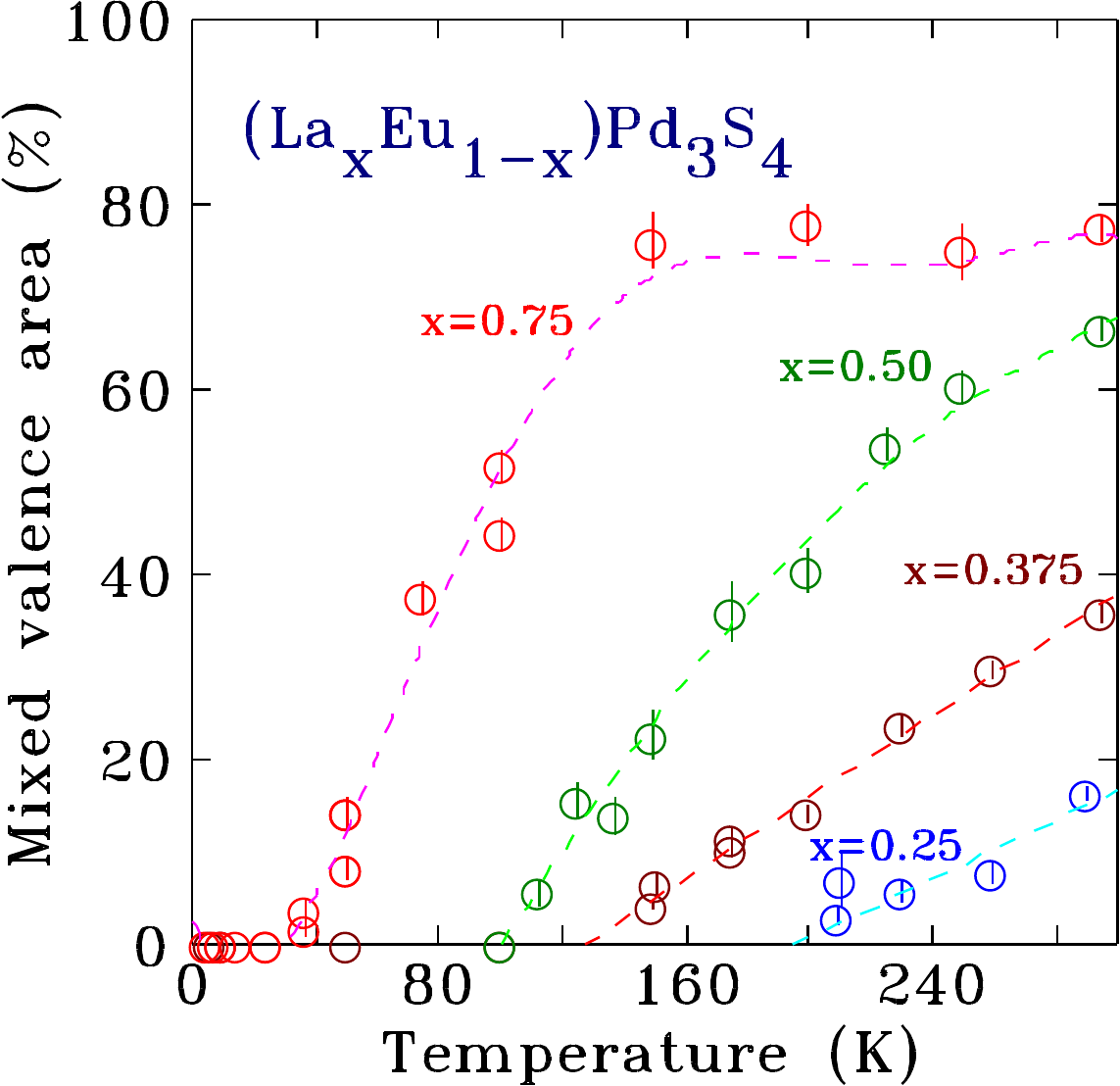}
\caption{Temperature dependence of the fitted area of the intermediate valence
feature for $\rm La_xEu_{1-x}Pd_3S_4$ showing both the greater extent and
earlier onset associated with increasing lanthanum substitution. Dashed lines are
guides to the eye.}
\label{fig:La-intermediate}
\end{figure}
%%%%%%%%%%%%%%%%%%%

We have presented low-temperature M\"ossbauer data for the yttrium- (5~K) and
lanthanum- (10~K) substituted series in order to minimise the impacts of the
differing Debye temperatures of the two valence components in the spectra.
Figures~\ref{fig:Y-RT-spectra} and \ref{fig:La-RT-spectra} present the room
temperature M\"ossbauer spectra for the $\rm Y_xEu_{1-x}Pd_3S_4$ and $\rm
La_xEu_{1-x}Pd_3S_4$ series. As can be seen, the differences between the room
temperature and 5~K yttrium data are minimal and certainly not qualitative. 
However, for the $\rm La_xEu_{1-x}Pd_3S_4$ series, the changes in the
spectra on warming go far beyond the simple, and expected, slight reduction in
the apparent Eu$^{2+}$ fraction with increasing temperature. As is clear from
the RT spectra of $\rm La_xEu_{1-x}Pd_3S_4$ shown in
Fig.~\ref{fig:La-RT-spectra}, an entirely new feature appears in the spectra and
eventually dominates the pattern by x=0.875. This feature is a line located
midway beween the lines associated with Eu$^{2+}$ and Eu$^{3+}$. As
Fig.~\ref{fig:La-RT-Relax_IS} shows,
the position of this line is essentially composition independent, appearing for
all compositions 0.25$\leq$x$<$1.00, and lying almost
precisely midway between the Eu$^{2+}$ and Eu$^{3+}$ lines, leading us to
identify it as being due to intermediate-valence europium: ``Eu$^{2.5+}$''.
In many other intermediate-valence europium compounds such as $\rm EuCu_2Si_2$
\cite{bauminger1053,palenzona199} and $\rm EuPd_2Si_2$ \cite{abd-elmeguid345},
where the electron exchange is between the europium and the conduction band,
\cite{rohler65,hossain014422} the isomer shift of the intermediate valence
component is strongly temperature dependent \cite{bauminger1053} as electrons
initially associated with the Eu$^{2+}$ ions spend more time in the conduction
band so that the average europium valence (and hence isomer shift) changes. That is not
the case here as the electron exchange is between Eu$^{2+}$ and Eu$^{3+}$ ions,
and as long as the residence time on each ion is roughly the same, then the
time-averaged valence is constant, even if the residence time (or equivalently,
the hopping rate) changes.

\begin{figure}
\includegraphics[width=7cm]{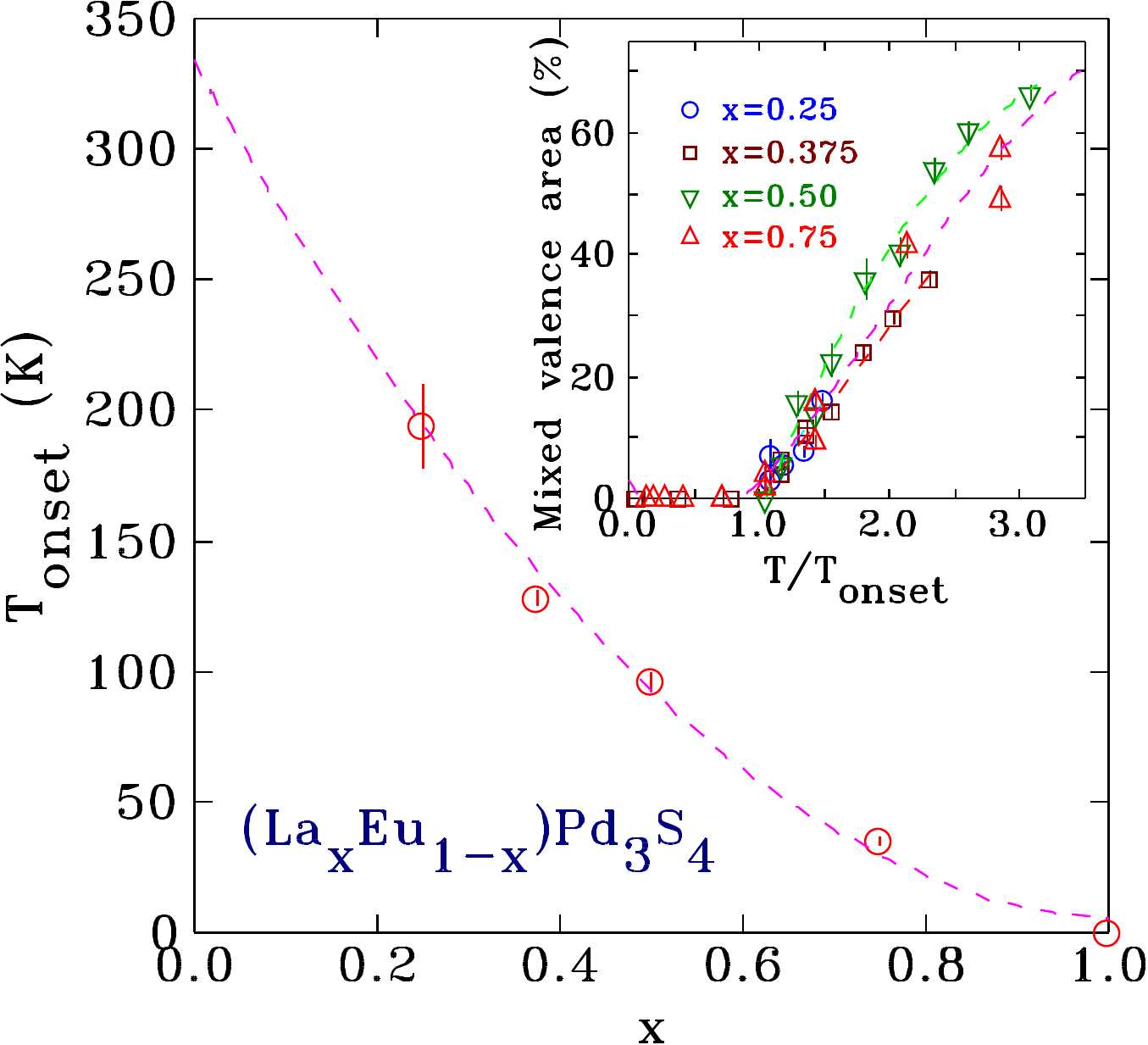}
\caption{Composition dependence of the temperature at which the intermediate
valence feature first appears in the $^{151}$Eu M\"ossbauer spectra of $\rm
La_xEu_{1-x}Pd_3S_4$ (see Fig.~\ref{fig:La-intermediate}). The dashed line is a
guide to the eye showing an extrapolation to x=0 that suggests a similar
transition for pure $\rm EuPd_3S_4$ near 340~K. The inset shows how the
composition and temperature dependences from Fig.~\ref{fig:La-intermediate}
collapse when scaled to T$_{onset}$.}
\label{fig:La-T-onset}
\end{figure}

The area of the intermediate valence feature is strongly dependent on both
temperature and composition (Fig.~\ref{fig:La-intermediate}).
At base temperature the area of the intermediate valence feature is zero, but as
temperature increases beyond T$_{onset}$ it starts to increase, with T$_{onset}$
decreasing with increasing x, {\it i.e.} increasing with increasing Eu concentration.
Fig~\ref{fig:La-T-onset} plots T$_{onset}$ versus x; the extrapolation of the
data to x=0, {\it i.e.} pure $\rm EuPd_3S_4$, suggests that T$_{onset}$ should
occur near T$\sim$340~K. The inset to Fig~\ref{fig:La-T-onset} shows the
intermediate valence area data presented in Fig.~\ref{fig:La-intermediate},
plotted versus effective temperature, T/T$_{onset}$. The data scale well and
this suggests that temperatures well above 340~K will need to be measures for
pure $\rm EuPd_3S_4$ in order to see the intermediate valence signal.
The progressive decrease in T$_{onset}$ apparent in Fig~\ref{fig:La-T-onset} for
the La-substituted series suggests that lanthanum substitution reduces the
barrier for electron hopping in $\rm La_xEu_{1-x}Pd_3S_4$. 

%%%%%%%%%%%%%%%%%%%
%%%%%%%%%%%%%%%%%%%

\begin{figure}
\includegraphics[width=7cm]{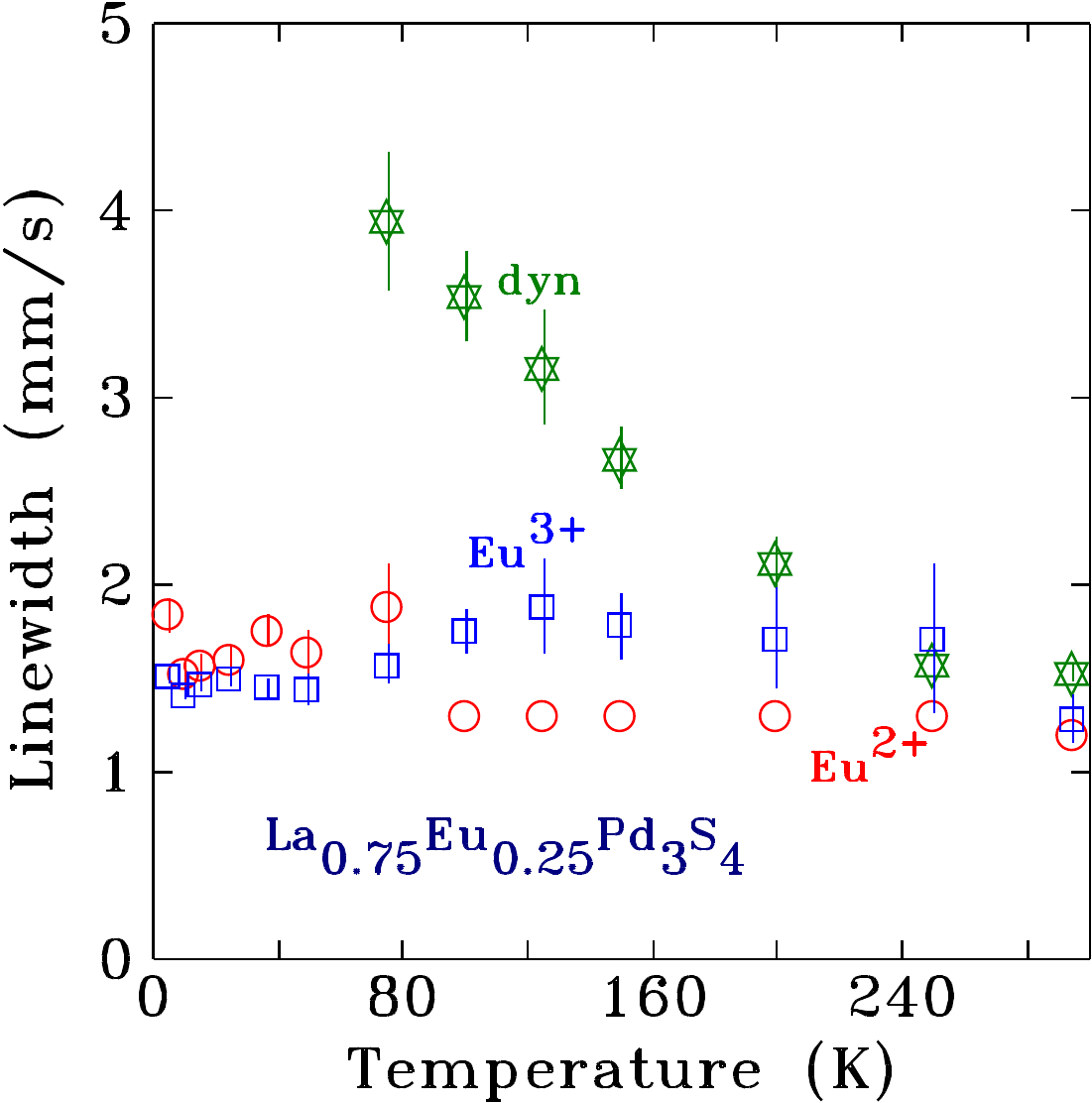}
\caption{Temperature dependence of the spectral linewidths for the three
components seen in the $^{151}$Eu M\"ossbauer spectra of $\rm
La_{0.75}Eu_{0.25}Pd_3S_4$. The Eu$^{2+}$ and Eu$^{3+}$ features have constant
linewidths, as expected. However, the dynamic feature associated with the
appearance of the  Eu$^{2.5+}$ component clearly sharpens with increasing
temperature reflecting motional narrowing associated with an increasing
relaxation rate. }
\label{fig:La-LW}
\end{figure}

Finally, although the position of the Eu$^{2.5+}$ line
appears to be both composition and temperature independent, its width is not. As
is clear from Fig.~\ref{fig:La-LW}, the Eu$^{2.5+}$ line becomes progressively
sharper with increasing temperature while the widths of the Eu$^{2+}$ and
Eu$^{3+}$ lines remain unchanged. This sharpening is characteristic of motional
narrowing.

Thermally driven conversion of Eu$^{2+}$ and Eu$^{3+}$ to an average Eu$^{2.5+}$
is consistent with rapid electron hopping between the two species. Clearly, the
two valence states are equally stable in $\rm EuPd_3S_4$ and in the absence of
evidence for charge ordering, it would appear that the spatial distribution of
the two species on the 2a site is essentially random. If a given site is
randomly $2+$ or $3+$, there would appear to be no reason to expect that
selection to be time-independent, and given sufficient thermal energy an
electron might be expected to jump from a Eu$^{2+}$
site to a neighbouring Eu$^{3+}$ site, exchanging the valence of the two sites
in the process. If the residence time of the electron is much longer than
the characteristic measuring time used to detect the valence ($\sim$10--100~ns
for M\"ossbauer spectroscopy) then the system appears static and two distinct
valence states are seen. If the residence time is much shorter, then an averaged
valence is seen. Intermediate residence times (``slow relaxation'') can lead to
more complex behaviour in the spectrum, but in the current case where the two
components are single lines with no significant quadrupole or magnetic
splittings, slow relaxation simply leads to a broad dynamic component. Raising
the temperature generally leads to faster hopping and a shorter residence time
and the system evolves from fully static, through the slow relaxation regime, to
dynamically averaged (or motionally narrowed). As can be seen for x=0.75 in
Fig.~\ref{fig:La-intermediate}, a weak broad line appears, it becomes stronger
and sharper (Fig.~\ref{fig:La-LW}) as more europium sites become dynamic and
motional narrowing occurs, finally developing into a natural-width line as the
hopping rate increases and moves up out of the M\"ossbauer window.

%%%%%%%%%%%%%%%%%%%
%%%%%%%%%%%%%%%%%%%

It is likely that the Eu$^{2+}$/Eu$^{3+}$ dynamics and the development of intermediate
valence europium are the cause of the difference between the estimate of percentage
of Eu$^{2+}$ between the M\"ossbauer, M(H) and M(T) data for
the lanthanum substituted samples shown in Fig.~\ref{fig:lattice}. The
$^{151}$Eu M\"ossbauer and the saturation magnetisation measuerements were made at
10~K and 1.8~K respectively, well below T$_{onset}$,
and the valence ratios derived from them are in agreement. However,
the Curie-Weiss analysis relies on measurements made at much higher
temperature ranges where the intermediate valence behaviour is both pronounced and
temperature dependent. Indeed, as the intermediate valence fraction becomes
larger (Fig.~\ref{fig:La-intermediate}), the Curie-Weiss-derived Eu$^{2+}$
fraction deviates more strongly from the M\"ossbauer and saturation
magnetisation values (Fig.~\ref{fig:lattice}). A simple Curie-Weiss analysis of
temperature-dependent susceptibility is not valid where the valence and hence
europium moment is strongly temperature dependent.

Finally, we note that if the same intermediate valence behaviour is present in
$\rm Y_xEu_{1-x}Pd_3S_4$ at RT it accounts for less than 10\% of the total area
(Fig.~\ref{fig:Y-RT-spectra}). The difference between the yttrium- and
lanthanum- substituted series may reflect stabilisation of Eu$^{2+}$ by
yttrium substitution, but there is also a very rapid loss of Eu$^{3+}$ as
yttrium is added, so as we approach the substitution levels where the
Eu$^{2.5+}$ component dominated in $\rm La_xEu_{1-x}Pd_3S_4$, there is almost no
Eu$^{3+}$ left in $\rm Y_xEu_{1-x}Pd_3S_4$ to exchange electrons with the
majority Eu$^{2+}$, so the absence of intermediate valence europium may simply
be the result of limited supply.

\subsection{Magnetic ordering}

\begin{figure}
\includegraphics[width=7cm]{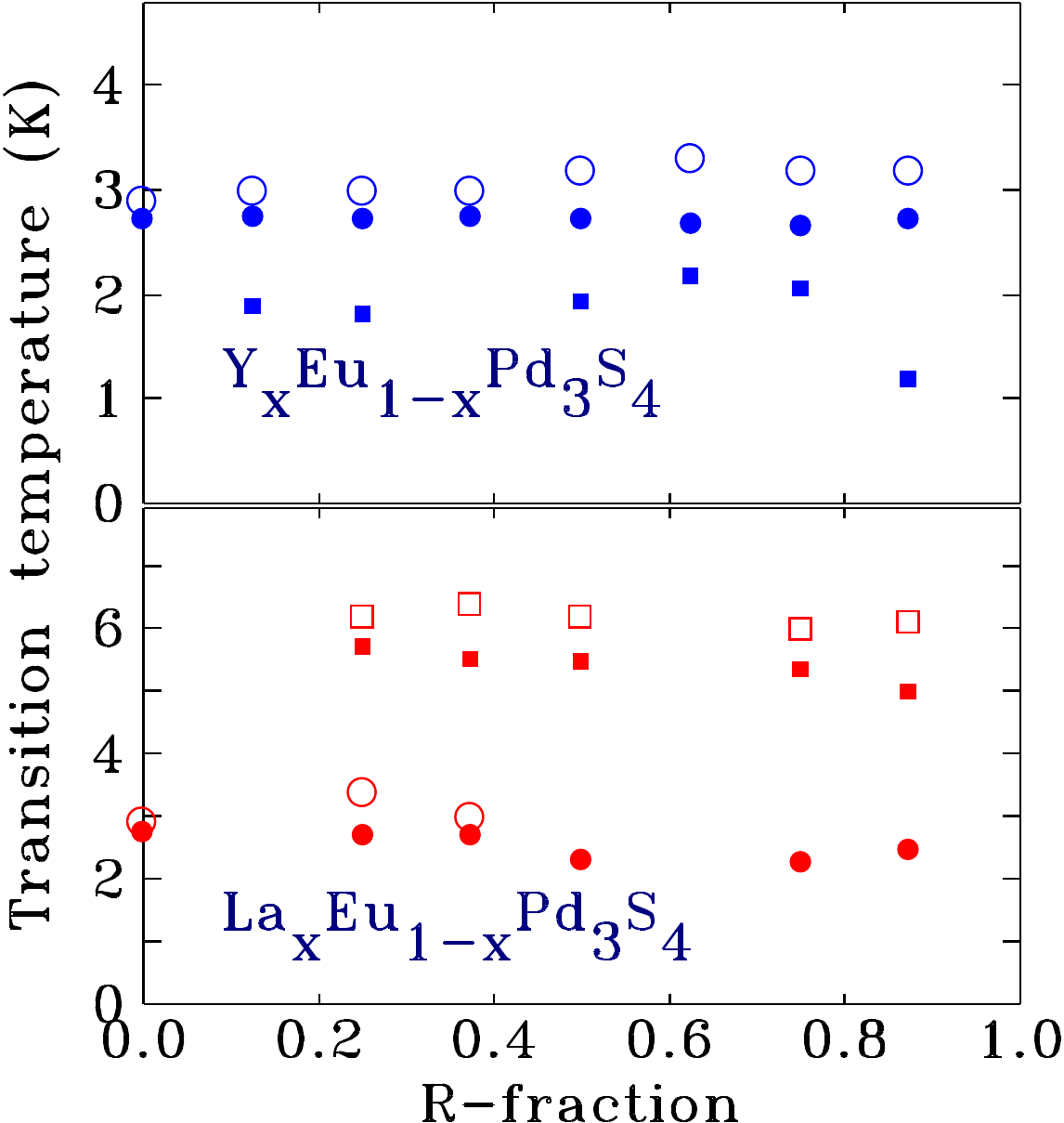}
\caption{Estimated transition temperatures for (top) $\rm Y_xEu_{1-x}Pd_3S_4$
and (bottom) $\rm La_xEu_{1-x}Pd_3S_4$. Open symbols show estimates derived from
dc susceptibility while solid symbols are based on C$_p$ data. }
\label{fig:Transitions}
\end{figure}

Although the Eu$^{2+}$ fraction does increase as yttrium is added, there is a
steady decline in the net moment-bearing rare-earth content of the $\rm
Y_xEu_{1-x}Pd_3S_4$ series with increasing x. This decline is even stronger in
the $\rm La_xEu_{1-x}Pd_3S_4$ series as the Eu$^{2+}$ fraction is either
constant (up to x$\sim$0.5) or decreases with increasing x. Despite these
monotonic reductions in the moment-bearing fraction, Fig.~\ref{fig:Transitions}
shows that both series continue to exhibit some form of magnetic order, at least
as far as x=0.875. Furthermore, it is apparent from both the high-field
magnetisation curves (Figs.~\ref{fig:Y-magn} and \ref{fig:La-magn}) and
low-field dc susceptibility (Figs.~ \ref{fig:Y-dcChi} and \ref{fig:La-dcChi} in
the appendix) that both series develop some ferromagnetic character to their
ordering. Heat capacity measurements (Figs.~\ref{fig:Y-Cp} and \ref{fig:La-Cp}
in the appendix) confirm the persistence of magnetic ordering. With increasing
dilution, the signature of the ordering becomes progressively weaker, with a
softer onset, but it is never lost.

%%%%%%%%%%%%%%%%%%%
%%%%%%%%%%%%%%%%%%%

For $\rm Y_xEu_{1-x}Pd_3S_4$ the onset of magnetic order appears to be largely
insensitive to the level of substitution, remaining near 3~K for all x.
However the Y-substituted samples also exhibit a secondary feature around 2~K in
the heat capacity data, and it too is largely independent of the composition.

For $\rm La_xEu_{1-x}Pd_3S_4$, heat capacity (Fig.~\ref{fig:La-Cp} in the
appendix) and dc susceptibility (Fig.~\ref{fig:La-dcChi} in the appendix)
measurements show that the ordering behaviour is more
complex in the lanthanum substituted series. Whereas the $\sim$3~K feature from the parent
compound persists for all finite x, but becomes progressively less distinct,
a new feature develops at $\sim$6~K as lanthanum is added. This is despite
the steady dilution of the moment bearing component: by $x=$0.875 only $\sim$4\% of the
rare earth atoms are moment-bearing Eu$^{2+}$, yet the 6~K magnetic feature is still
apparent in both heat capacity and susceptibility. $^{151}$Eu M\"ossbauer spectroscopy
confirms that the 6~K feature at x=0.75 is indeed magnetic and associated with the
Eu$^{2+}$ in the material. Although the $\sim$5~K limit of the closed-cycle fridge
was not low enough to yield a fully resolved magnetic splitting,
Fig.~\ref{fig:La-width} (in the appendix) shows that it is possible to observe the initial line
broadening that occurs as order develops. Fitting the broad Eu$^{2+}$ line at
5.3~K yields a hyperfine field of 11(1)~T.
Fig.~\ref{fig:Transitions} summarises the composition dependence of the transition
temperatures observed for $\rm (Y,La)_xEu_{1-x}Pd_3S_4$.

The form of C$_p$(T) for $\rm La_{0.25}Eu_{0.75}Pd_3S_4$, and to a lesser extent
$\rm La_{0.375}Eu_{0.625}Pd_3S_4$, might lead one to suspect that either an
impurity phase is present, or that the sample has perhaps separated into La-rich and
La-poor regions, yielding distinct $\sim$6~K and $\sim$3~K transitions
respectively. The magnitude of the C$_p$ signal would require a rather
significant impurity content to account for it. None was detected by x-ray
diffraction, as noted earlier. Furthermore, the change in lattice parameter with
x in the La-substituted series is quite strong (see Fig.~\ref{fig:lattice}) so any
phase separation would yield broadened, or more likely split, diffraction lines.
These were not seen in any sample studied here. Finally, the $\rm NaPt_3O_4$
form of $\rm RPd_3S_4$ appears to be stable across the entire rare-earth series
and for both the Y- and La- substituted series studied here, giving us no reason
to expect phase separation. The monotonic composition and temperature evolution
of the intermediate valence component apparent in Figs.~\ref{fig:La-RT-spectra}
and \ref{fig:La-intermediate} also argues for homogeneous, rather than phase
separated samples. The $^{151}$Eu M\"ossbauer spectra in
Fig.~\ref{fig:La25-pumped-spec} (in the appendix) reveal a progressive
development of magnetic order on cooling: The Eu$^{2+}$ component near
$-$11~mm/s first broadens below 6~K, then a full magnetic pattern appears (more
apparent on the low-velocity side of the Eu$^{2+}$ component) on further
cooling. Even at 1.8~K the magnetic component is extremely broad (compare the
1.8~K spectrum in Fig.~\ref{fig:La25-pumped-spec} (in the appendix) with that of
$\rm EuPd_3S_4$ at the same temperature in Fig.~\ref{fig:Eu-spectra}). A simple
parametrisation of the spectra was obtained by using three components to fit the
Eu$^{2+}$ contribution: zero field, low-field and high-field. Fewer components
gave a poor fit while using more led to instabilities. Following the areas of
the magnetic and non-magnetic Eu$^{2+}$ components
(Fig.~\ref{fig:La25-pumped-area} in the appendix) shows a gradual shift from
non-magnetic to magnetic starting below 6~K with no apparent break near 3~K. The
average magnetic field shown in Fig.~\ref{fig:La25-pumped-field} (in the
appendix) shows similar behaviour, again with no break near 3~K. We conclude
that although the Eu$^{2+}$ experiences a very broad range of magnetic
environments, there is no evidence for phase separation.

\section{Discussion}

Given that the $\rm RPd_3S_4$ compound series exists for most, if not all of the
trivalent rare earths (R) and yttrium \cite{keszler2369,
wakeshima1, wakeshima226, bonville263}, but not, apparently, for the divalent
alkaline earths, it is remarkable that $\rm EuPd_3S_4$ not
only accommodates such a large fraction of divalent europium, but that the fraction
of Eu$^{2+}$ is so robust against substitution by other trivalent rare earths:
Substituting yttrium actually promotes the divalent state, whereas substituting
lanthanum has little impact on the europium valence until at least half of the
europium has been replaced.

It is clear that band filling or average valence electron count does not control
the Eu$^{2+}$:Eu$^{3+}$ ratio in $\rm R_xEu_{1-x}Pd_3S_4$ as both yttrium and
lanthanum are trivalent yet act on the Eu$^{2+}$:Eu$^{3+}$ ratio in opposite
directions. Similarly, chemical pressure arguments fail as lattice compression
using yttrium substitution favours the larger Eu$^{2+}$ ion, and lattice
expansion using lanthanum leads to more of the smaller Eu$^{3+}$ ion -- the
reverse of what would be predicted.

We are left with preservation of the average rare-earth size as the dominant
factor controlling the Eu$^{2+}$:Eu$^{3+}$ valence ratio. Since $r_{ionic} \rm
(Y^{3+}$) is both smaller than $r_{ionic} \rm (Eu^{3+}$) and much smaller than
$r_{ionic} \rm (Eu^{2+}$), converting Eu$^{3+}$ into the larger Eu$^{2+}$
as yttrium is added acts to maintain the average rare-earth size. By the same
token, $r_{ionic} \rm (La^{3+}$) lies between those of Eu$^{2+}$ and Eu$^{3+}$,
so lanthanum substitution initially has very little effect on the Eu$^{2+}$:Eu$^{3+}$
ratio. Only after half of the europium has been replaced do we see a shift in
the balance towards Eu$^{3+}$.

Hydrostatic pressure would provide a complementary window on how pure changes
in volume affect the Eu$^{2+}$:Eu$^{3+}$ ratio, and synchrotron M\"ossbauer
spectroscopy under pressure would be a useful extension of this study.

Remarkably, La-substitution also appears to reduce the barrier for electron hopping
between the Eu$^{2+}$ and Eu$^{3+}$ ions in $\rm La_xEu_{1-x}Pd_3S_4$, and
a temperature dependent intermediate valence component develops at
progressively lower temperatures with increasing lanthanum content.
Extrapolating the composition trend in the onset temperature leads to the
prediction that electron hopping should start around 340~K in un-doped $\rm
EuPd_3S_4$. The much
weaker dynamic behaviour in the Y-substituted series may reflect a stronger
bias towards Eu$^{2+}$ in $\rm Y_xEu_{1-x}Pd_3S_4$ which may in turn
stabilise the valence of the two europium components. Alternatively, the much
lower availabilty of Eu$^{3+}$ for electron exchange may be the limiting factor.

The evolution of low temperature magnetism, as seen by M\"ossbauer spectroscopy,
heat capacity and low field magnetic susceptibility also differs 
between yttrium and lanthanum. Remarkably, suppression of the ordering by the dilution
of the magnetic Eu$^{2+}$ is not the primary impact in either series.
Magnetic order persists in $\rm Y_xEu_{1-x}Pd_3S_4$ even as far
as x=0.875, with the onset temperature essentially unchanged, despite only about
12\% of the europium sites being occupied by a moment-bearing ion
(Fig.~\ref{fig:Transitions}). However the
behaviour of the La-substituted series is perhaps more surprising. Not only does
the magnetic order also persist to x=0.875 (where the moment-bearing fraction is
only $\sim$4\% because of the much lower Eu$^{2+}$:Eu$^{3+}$ ratio in the
La-substituted series), but the onset temperature jumps from $\sim$3~K
to $\sim$6~K for low x and remains near 6~K at least as far as x=0.875
(Fig.~\ref{fig:Transitions}). We emphasise that our data do not provide
access to the specific magnetic structures of the Y- and La- substituted
materials. The behaviour of $\chi$(T) and C$_p$(T) in Fig.~\ref{fig:Eu-Cp-chi}
strongly suggests that the parent compound orders antiferromagnetically,
however it is unlikely that this survives the
substitutional disorder introduced when we replace the europium. The form of
$\chi$(T) clearly changes in both series (Figs.~\ref{fig:Y-dcChi} and
~\ref{fig:La-dcChi} in the appendix) and the magnetic order appears
to develop at least some ferromagnetic character, although it is probably
dominated by short ranged order.
The extremely broad hyperfine field distributions and the
presence of significant zero field components well below the onset of magnetic
order in both $\rm La_{0.25}Eu_{0.75}Pd_3S_4$ and $\rm Y_{0.25}Eu_{0.75}Pd_3S_4$
(Fig.~\ref{fig:1p8Kspectra}) point to inhomogeneous ordering in both of these magnetically
diluted materials.

\section{Conclusions}

The impact of yttrium and lanthanum substitution on mixed-valence $\rm EuPd_3S_4$ has
been studied using $^{151}$Eu M\"ossbauer spectroscopy, bulk magnetisation and
heat capacity measurements. Average valence electron count clearly does not control the
europium valence distribution as trivalent yttrium and lanthanum substitutions have
opposite effects. Similarly, chemical pressure arguments fail as lattice
compression using yttrium substitution favours the larger Eu$^{2+}$ ion, with
the reverse effect seen with lanthanum. It appears that preservation of the
average rare-earth size is the dominant factor controlling the
Eu$^{2+}$:Eu$^{3+}$ valence ratio.

Lanthanum substitution was also found to promote electron hopping between the
Eu$^{2+}$ and Eu$^{3+}$ ions, leading to the formation of intermediate valence
europium. Increasing lanthanum substitution both increases the amount of
intermediate valence europium seen at ambient temperatures and reduces the
temperature at which hopping starts to appear. An onset temperature of
$\sim$340~K is predicted for undoped $\rm EuPd_3S_4$.

Inhomogeneous magnetic order is seen at all levels of substitution, with the
onset temperature in the Y-substituted series being essentially unchanged at
3~K, while for the La-substituted series the onset temperature is seen to
increase to 6~K despite the low concentration of magnetic ions.

\begin{acknowledgments}

Financial support for this work was provided by
Fonds Qu\'eb\'ecois de la Recherche sur la Nature et les Technologies,
and the Natural Sciences and Engineering Research Council (NSERC) Canada.

Work at the Ames Laboratory was supported by the U.S. Department of Energy,
Office of Science, Basic Energy Sciences, Materials Sciences and Engineering
Division. The Ames Laboratory is operated for the U.S. Department of Energy by
Iowa State University under contract No. DE-AC02-07CH11358.

BK is supported by the Center for the Advancement of Topological Semimetals
(CATS), an Energy Frontier Research Center funded by the US DOE, Office of Basic
Energy Sciences.

Much of this work was carried out while DHR was on sabbatical at Iowa State
University and their generous support during this visit is gratefully acknowledged.

\end{acknowledgments}

% \bibliography{eupd3s4.bib}

% ====================================================================================


% ====================================================================================

\appendix

\section{Additional figures and tables}

\renewcommand{\thefigure}{A.\arabic{figure}}
\setcounter{figure}{0}

\renewcommand{\thetable}{A.\arabic{table}}
\setcounter{table}{0} 

This appendix contains additional figures showing heat capacity and
susceptibility data for $\rm Y_xEu_{1-x}Pd_3S_4$ and $\rm La_xEu_{1-x}Pd_3S_4$.

We also present low temperature $^{151}$Eu M\"ossbauer data for
$\rm La_{0.5}Eu_{0.5}Pd_3S_4$ showing the onset of magnetic broadening near 6~K, and
spectra for $\rm La_{0.25}Eu_{0.75}Pd_3S_4$ showing the inhomogeneous magnetic
order that develops below the transition temperature. The temperature dependence of some fitting
parameters for the $\rm La_{0.25}Eu_{0.75}Pd_3S_4$ spectra is also shown.

The fitted lattice parameters and Eu$^{2+}$ fractions used to construct
Fig.~\ref{fig:lattice} in the main body of the text are also tabulated here.

%%%%%%%%%%%%%%%%%%%
\begin{figure*}
\includegraphics[width=13cm]{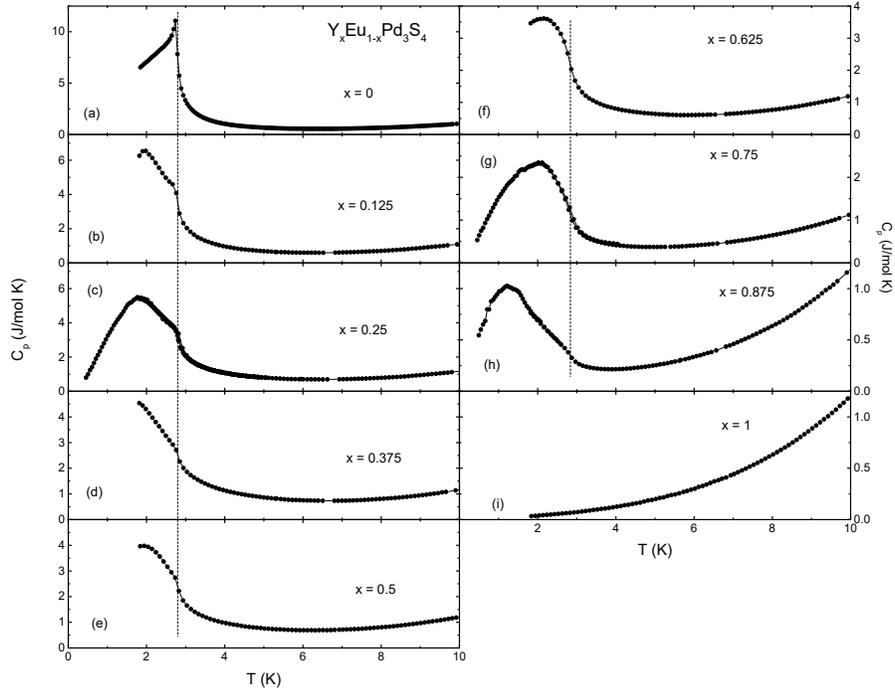}
\caption{Heat capacity measurements for $\rm Y_xEu_{1-x}Pd_3S_4$
showing the persistence of the magnetic transition near 3~K despite the
significant dilution of the magnetic Eu$^{2+}$. At x=1 there is no transition
observed as the system contains no magnetic species and so is paramagnetic.
The vertical dashed lines mark 3~K. }
\label{fig:Y-Cp}
\end{figure*}
%%%%%%%%%%%%%%%%%%%

%%%%%%%%%%%%%%%%%%%
\begin{figure*}
\includegraphics[width=13cm]{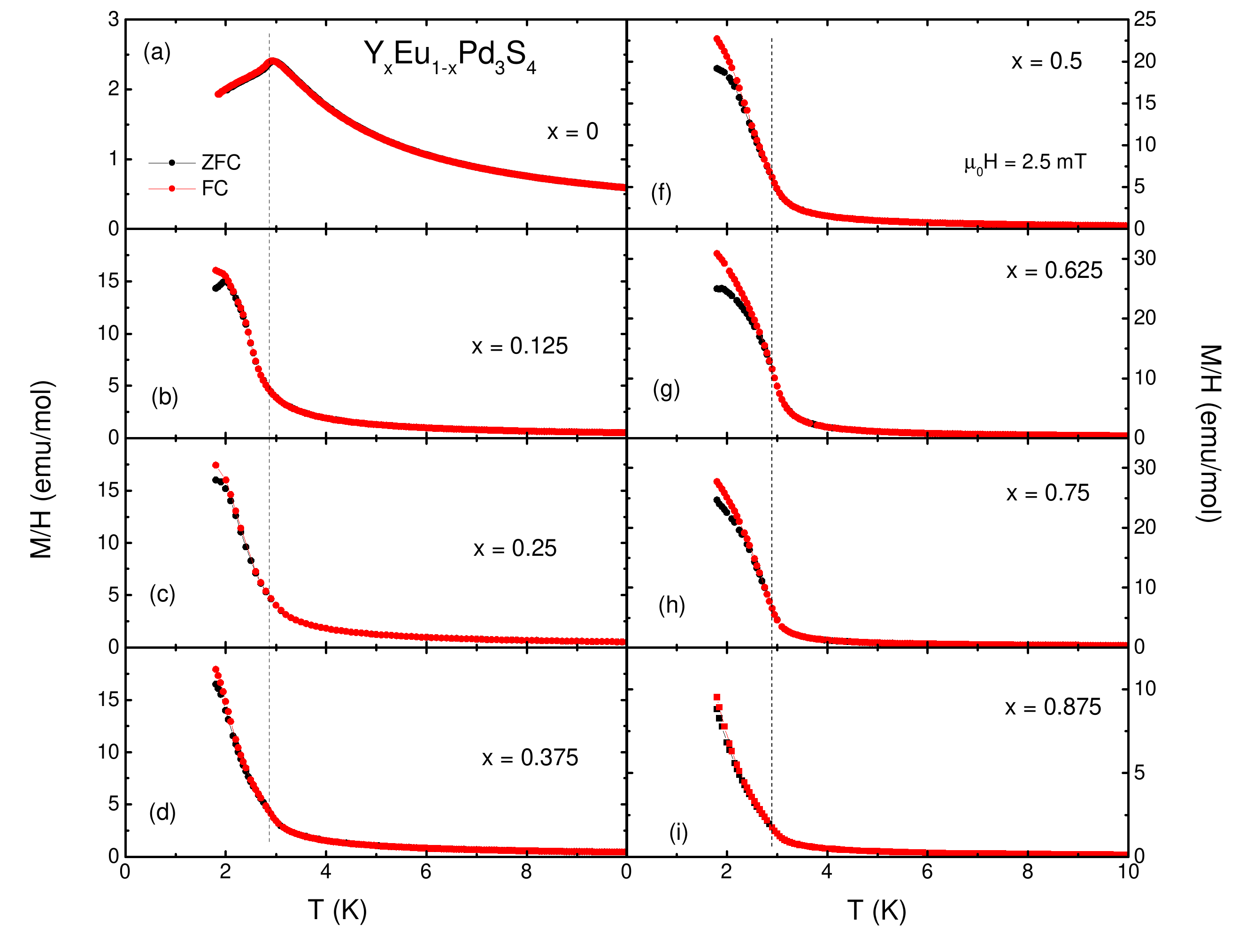}
\caption{Field-cooled (FC) and zero-field-cooled (ZFC) dc susceptibility for
$\rm Y_xEu_{1-x}Pd_3S_4$ confirming the persistence of the magnetic transition
seen near 3~K in the heat capacity data (Fig.~\ref{fig:Y-Cp}) despite the
significant dilution of the magnetic Eu$^{2+}$. The vertical dashed lines mark
3~K. }
\label{fig:Y-dcChi}
\end{figure*}
%%%%%%%%%%%%%%%%%%%

%%%%%%%%%%%%%%%%%%%
\begin{figure*}
\includegraphics[width=12.5cm]{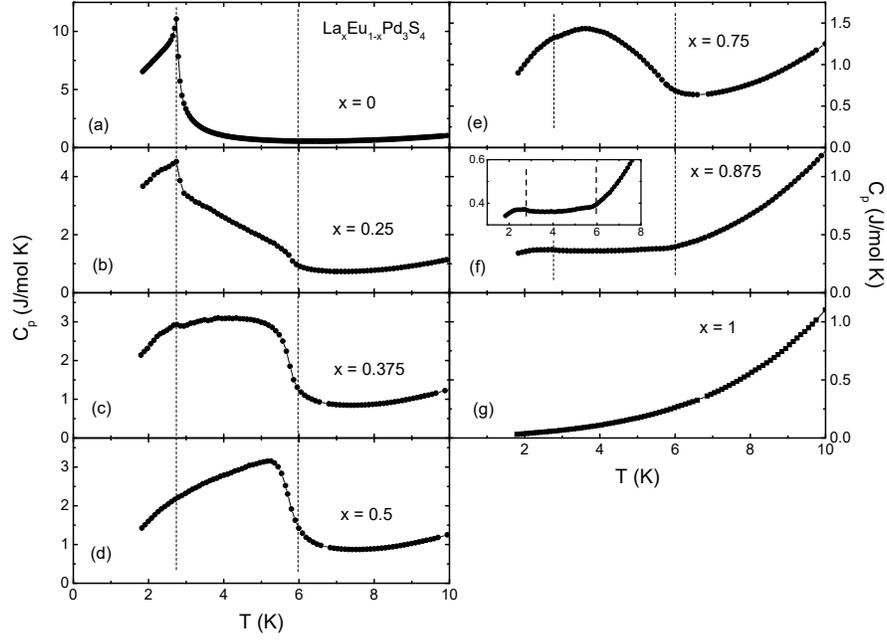}
\caption{Heat capacity measurements for $\rm La_xEu_{1-x}Pd_3S_4$ showing the
gradual replacement of the 3~K transition by a new feature at 6~K, despite the
significant dilution of the magnetic Eu$^{2+}$ component. The inset in (f) shows
the region between 2~K and 8~K on an expanded vertical scale. At x=1 there is no
transition observed as the system contains no magnetic species and so is
paramagnetic. The vertical dashed lines mark 3~K and 6~K.}
\label{fig:La-Cp}
\end{figure*}
%%%%%%%%%%%%%%%%%%%

%%%%%%%%%%%%%%%%%%%
\begin{figure*}
\includegraphics[width=12.5cm]{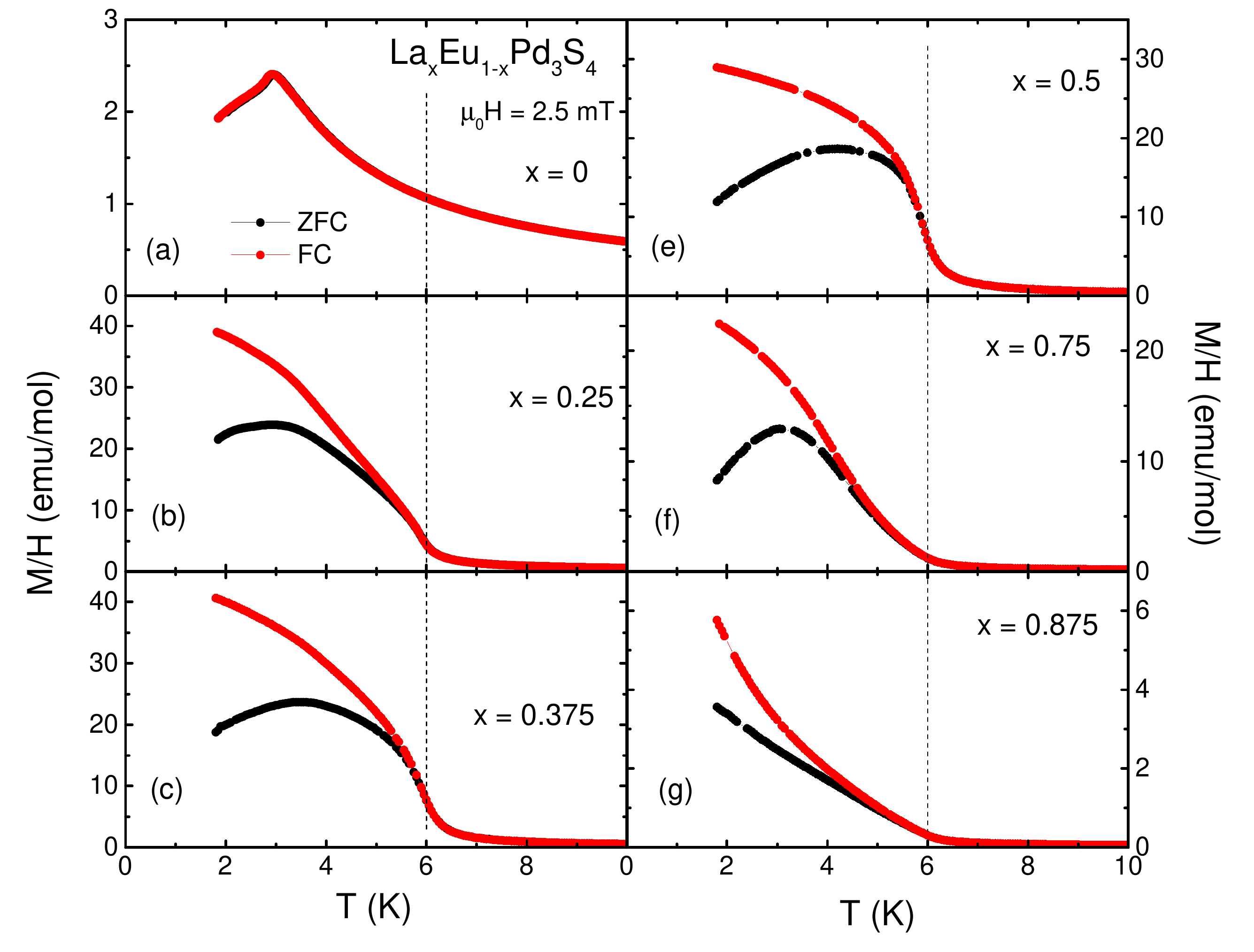}
\caption{Field-cooled (FC) and zero-field-cooled (ZFC) dc susceptibility for
$\rm La_xEu_{1-x}Pd_3S_4$ confirming the gradual replacement of the 3~K
transition by a new feature at 6~K,
seen in the heat capacity data (Fig.~\ref{fig:La-Cp}) despite the
significant dilution of the magnetic Eu$^{2+}$. The vertical dashed lines mark
3~K. }
\label{fig:La-dcChi}
\end{figure*}
%%%%%%%%%%%%%%%%%%%

%%%%%%%%%%%%%%%%%%%
\begin{figure}
\includegraphics[width=7cm]{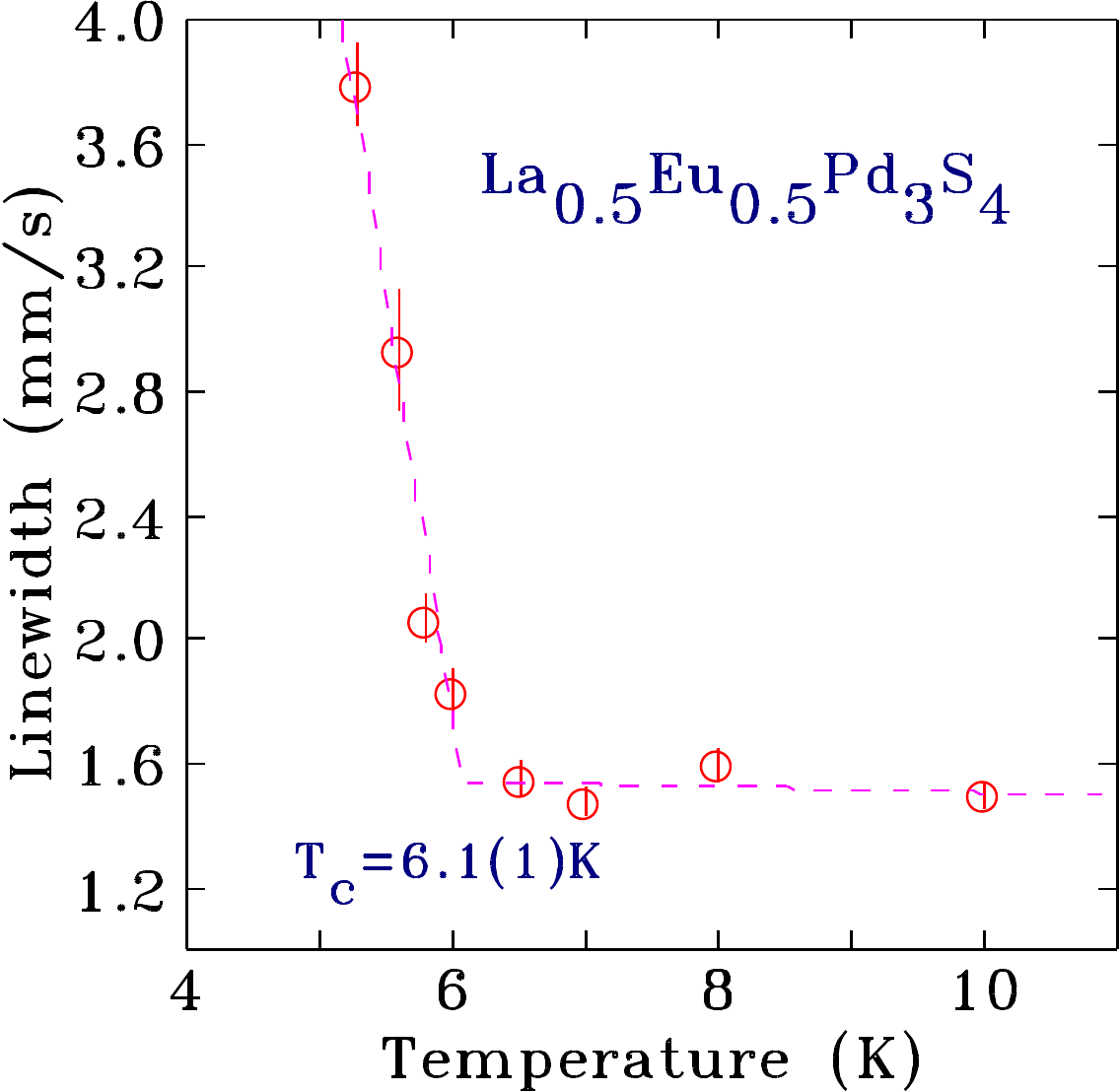}
\caption{Temperature dependence of the Eu$^{2+}$ line width in $\rm
La_{0.5}Eu_{0.5}Pd_3S_4$ showing the onset of ordering at 6.1(1)~K.}
\label{fig:La-width}
\end{figure}
%%%%%%%%%%%%%%%%%%%

%%%%%%%%%%%%%%%%%%%
\begin{figure}
\includegraphics[width=7cm]{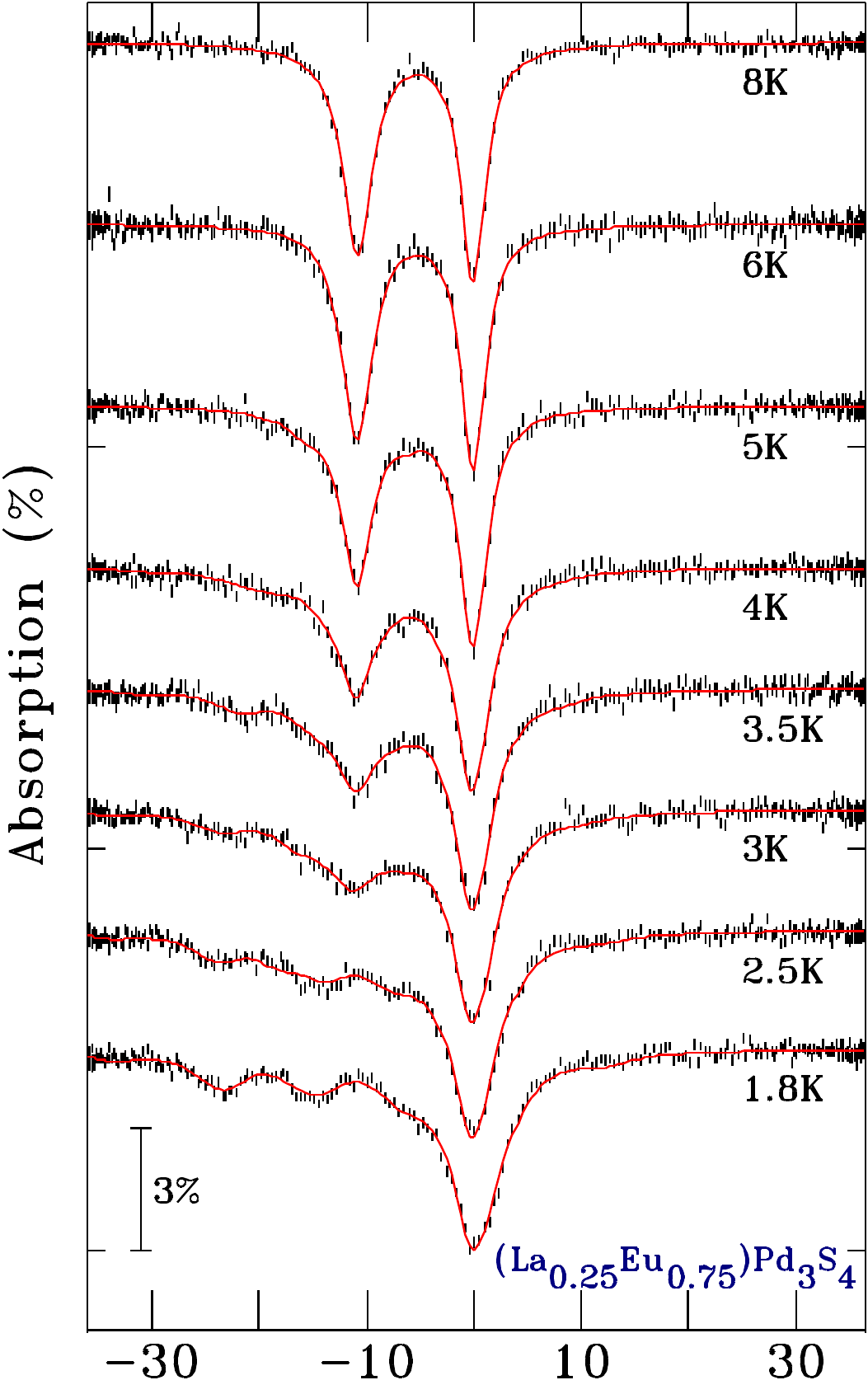}
\caption{$^{151}$Eu M\"ossbauer spectra of $\rm La_{0.25}Eu_{0.75}Pd_3S_4$
showing the progressive development of magnetic order in the Eu$^{2+}$
component. The solid lines are fits using a single line for the Eu$^{3+}$ line
near 0~mm/s, and three components to fit the Eu$^{2+}$: an unsplit line, and two
magnetic contributions with a ``small'' and ``large'' field used to reproduce
the gross behaviour of the system.}
\label{fig:La25-pumped-spec}
\end{figure} 
%%%%%%%%%%%%%%%%%%%

%%%%%%%%%%%%%%%%%%%
\begin{figure}
\includegraphics[width=7cm]{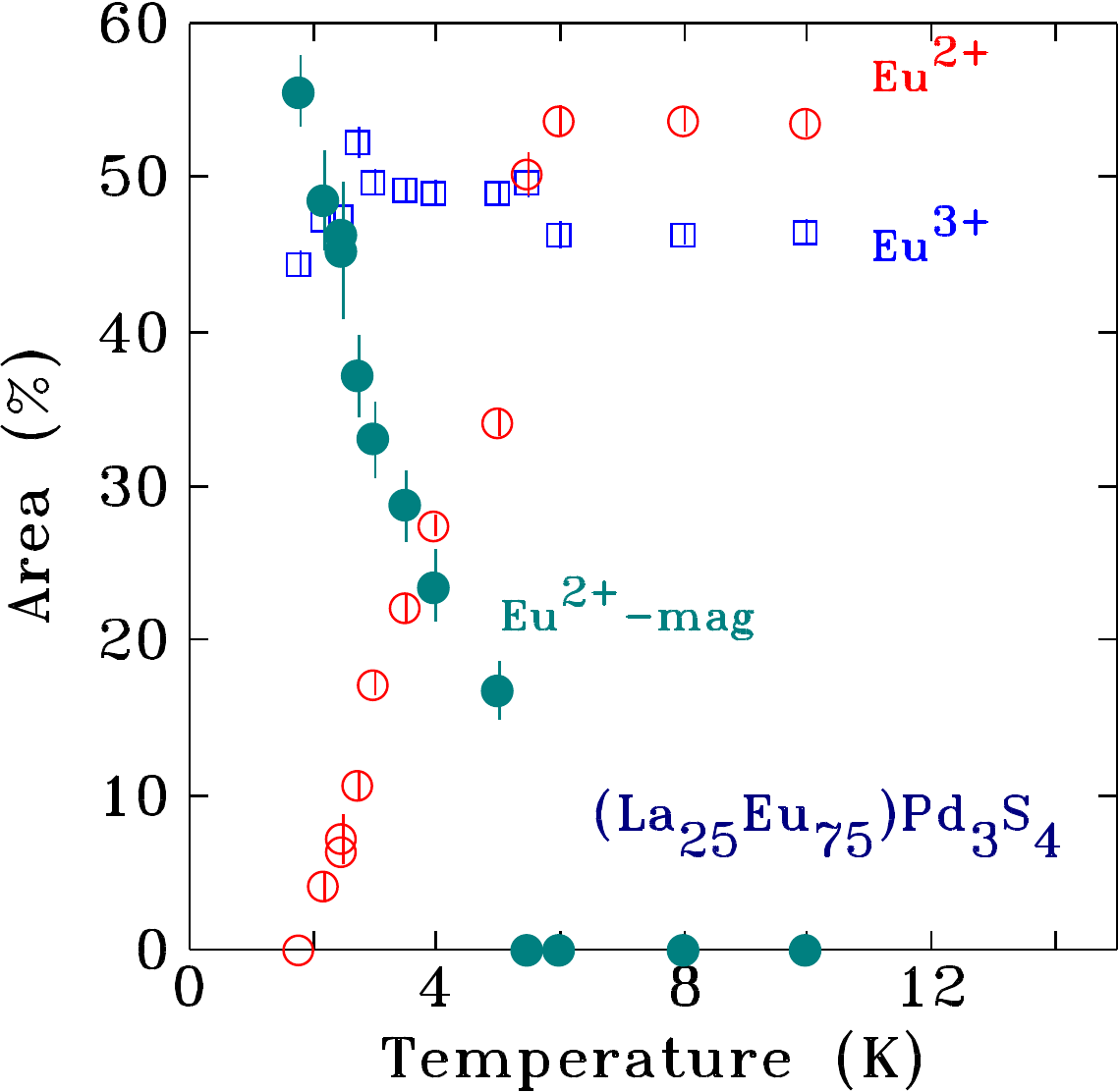}
\caption{Temperature dependence of the magnetic and non-magnetic Eu$^{2+}$
compoents in the $^{151}$Eu M\"ossbauer spectra of $\rm
La_{0.25}Eu_{0.75}Pd_3S_4$ showing the gradual onset of magnetic order below 6~K
but no evidence for a break at 3~K that might suggest the presence of a second
phase.}
\label{fig:La25-pumped-area}
\end{figure}
%%%%%%%%%%%%%%%%%%%

%%%%%%%%%%%%%%%%%%%
\begin{figure}
\includegraphics[width=7cm]{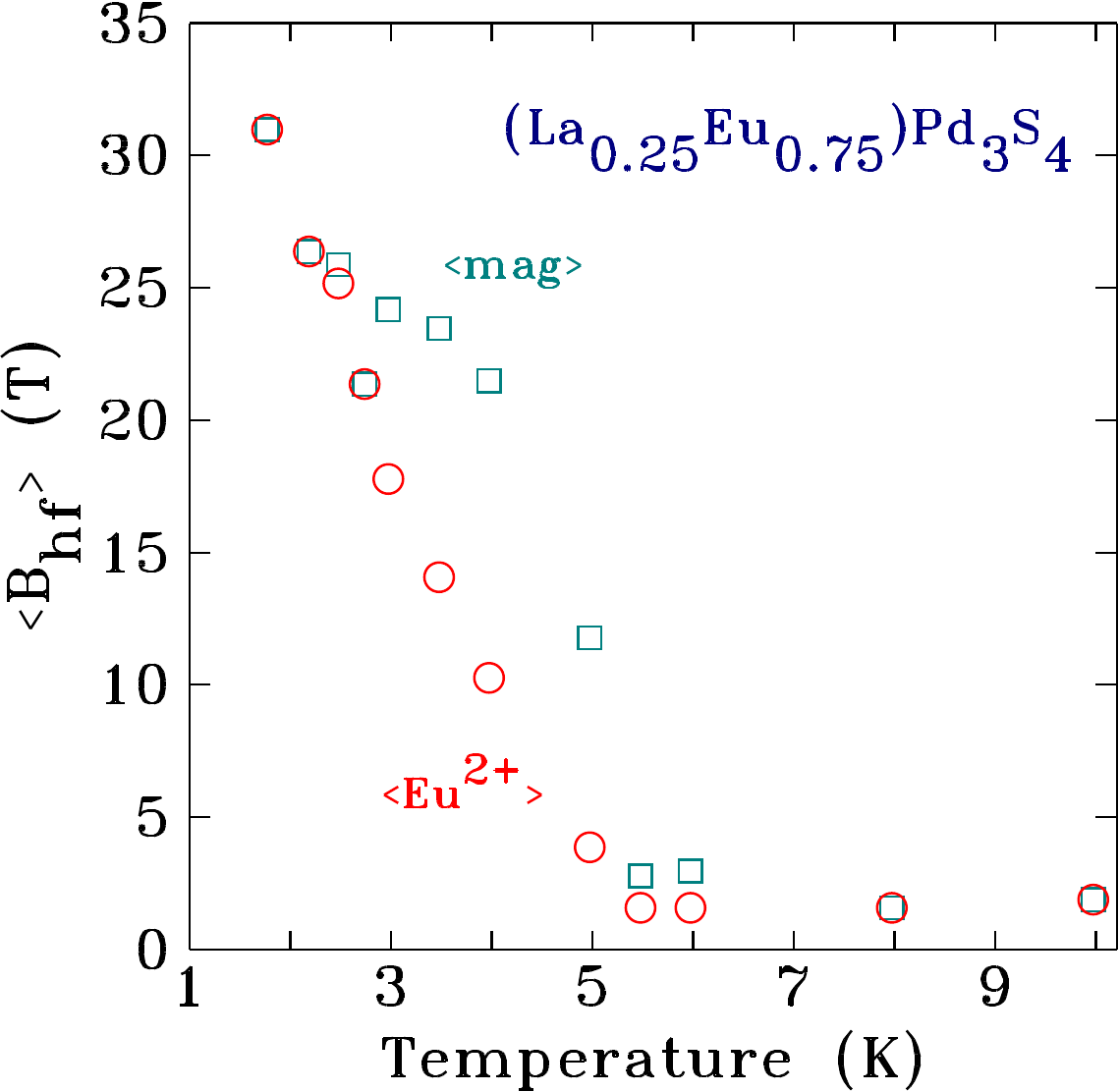}
\caption{Temperature dependence of the average Eu$^{2+}$ magnetic hyperfine field
(B$_{hf}$) in the $^{151}$Eu M\"ossbauer spectra of $\rm La_{0.25}Eu_{0.75}Pd_3S_4$
showing the gradual onset of magnetic order below 6~K. Two averages are shown:
(i) $\rm <Eu^{2+}>$ (blue circles) is calculated for all three Eu$^{2+}$ components
used in the fits; (ii) $\rm <mag>$ (red squares) is calculated omitting the
Eu$^{2+}$ component that exhibits no magnetic splitting. See text for details of
the model used to fit the spectra.}
\label{fig:La25-pumped-field}
\end{figure}
%%%%%%%%%%%%%%%%%%%

%%%%%%%%%%%%%%%%%%%
\begin{figure}
\includegraphics[width=7cm]{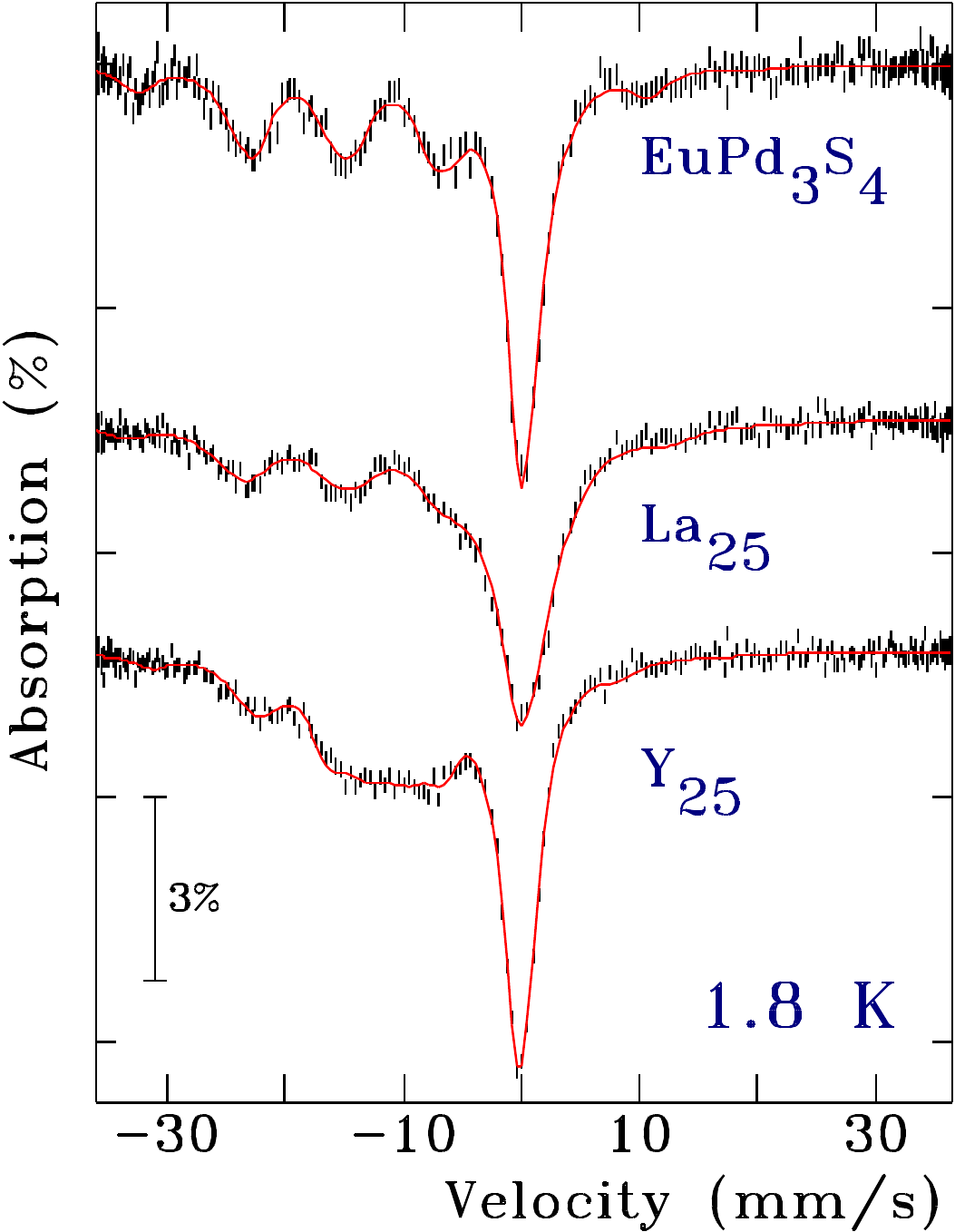}
\caption{$^{151}$Eu M\"ossbauer spectra of $\rm EuPd_3S_4$ (top),
$\rm La_{0.25}Eu_{0.75}Pd_3S_4$ (middle) and $\rm Y_{0.25}Eu_{0.75}Pd_3S_4$ 
(bottom) measured at 1.8~K showing that while the substituted materials do appear to
exhibit magnetic ordering of the Eu$^{2+}$ component, the lines are not sharp,
reflecting a broad distribution of hyperfine fields. }
\label{fig:1p8Kspectra}
\end{figure} 
%%%%%%%%%%%%%%%%%%%

\pagebreak

% \vspace*{2cm}

\begin{table}
\begin{tabular}{l|c|c|c|c}
\hline
\ \ x \ \  & \ \ \ \ \  a (\AA) \ \ \ \ \  & \multicolumn{3}{|c}{ Eu$^{2+}$
fraction (\%)} \\
& & \ \ M\"ossbauer \ \ & \ \ Curie-Weiss \ \ & M(H) \\
& & & \ \ $\pm$5\% \ \ & $\pm$5\% \\
 \hline
 & & & & \\
0.00 & 6.67858(8) & 53.8(5) & 57 & 49\\
0.125 & 6.67637(9) & 56.3(8) & 59 & 56 \\
0.25 & 6.67371(13) & 61.5(5) & 66 & 63 \\
0.375 & 6.66983(12) & 68.4(5) & 72 & 68 \\
0.500 & 6.66527(20) & 73.3(9) & 71 & 69 \\
0.625 & 6.66148(10) & 83.2(7) & 92 & 89 \\
0.750 & 6.65514(12) & 87.3(8) & 85 & 86 \\
0.875 & 6.64921(11) & 93(2) & 95 & 94 \\
1.00 & 6.64193(17) & -- & -- & -- \\
 & & & & \\
 \hline 
\end{tabular}
\caption{Measured lattice parameters and Eu$^{2+}$ fractions for the $\rm
Y_xEu_{1-x}Pd_3S_4$ compound series.}
\label{tab:yttrium}
\end{table}

\begin{table}
\begin{tabular}{l|c|c|c|c}
\hline
\ \ x \ \  & \ \ \ \ \  a (\AA) \ \ \ \ \  & \multicolumn{3}{|c}{ Eu$^{2+}$
fraction (\%)} \\
& & \ \ M\"ossbauer \ \ & \ \ Curie-Weiss \ \ & M(H) \\
& & & \ \ $\pm$5\% \ \ & $\pm$5\% \\
 \hline
 & & & & \\
0.00 & 6.67858(8) & 53.8(5) & 57 & 49 \\
0.25 & 6.69217(18) & 53.1(6) & 55 & 52 \\
0.375 & 6.70125(17) & 51.1(9) & 53 & 49 \\
0.50 & 6.70871(17) & 51.5(8) & 54 & 49 \\
0.625 & 6.71763(12) & 47(2) & 53 & 46 \\
0.75 & 6.72490(14) & 46(2) & 56 & 43 \\
0.8125 & 6.72981(12) & 39(1) & 57 & 40 \\
0.875 & 6.73316(14) & 32(1) & 61 & 36 \\
0.9375 & 6.73557(6) & 25(3) & 86 & 34 \\
1.00 & 6.73948(9) & -- & -- & -- \\
 & & & & \\
 \hline 
\end{tabular}
\caption{Measured lattice parameters and Eu$^{2+}$ fractions for the $\rm
La_xEu_{1-x}Pd_3S_4$ compound series.}
\label{tab:lanthanum}
\end{table}

\end{document}